\documentclass[acmengage]{acmart}

\AtBeginDocument{%
  \providecommand\BibTeX{{%
    \normalfont B\kern-0.5em{\scshape i\kern-0.25em b}\kern-0.8em\TeX}}}

\setcopyright{cc}
\setcctype[4.0]{by}
\copyrightyear{2026}
\acmYear{2026}
\acmDOI{10.1145/3772318.3791242}

\acmConference[CHI '26]{the
2026 CHI Conference on Human Factors in Computing Systems}{April 13–17, 2026}{Barcelona, Spain}
\acmISBN{979-8-4007-2278-3/26/04}

\usepackage{array}

\begin{document}

\title{"I Don't Trust Any Professional Research Tool": A Re-Imagination of Knowledge Production Workflows by, with, and for Blind and Low-Vision Researchers}

\author{Omar Khan}
\email{mkhan259@illinois.edu}
\orcid{0009-0005-3209-3525}
\affiliation{%
    \department{Siebel School of Computing and Data Science}
    \institution{University of Illinois Urbana-Champaign}
    \city{Urbana}
    \state{Illinois}
    \country{USA}
}

\author{JooYoung Seo}
\email{jseo1005@illinois.edu}
\orcid{0000-0002-4064-6012}
\affiliation{%
    \department{School of Information Sciences}
    \institution{University of Illinois Urbana-Champaign}
    \city{Champaign}
    \state{Illinois}
    \country{USA}
}

\renewcommand{\shortauthors}{Khan et al.}
\renewcommand{\shorttitle}{["I Don't Trust Any Professional Research Tool"]}

\begin{abstract}
Research touts universal participation through accessibility initiatives, yet blind and low-vision (BLV) researchers face systematic exclusion as visual representations dominate modern research workflows.
 To materialize inclusive processes, we, as BLV researchers, examined how our peers combat inaccessible infrastructures. 
Through an explanatory sequential mixed-methods approach, we conducted a cross-sectional, observational survey (n=57) and follow-up semi-structured interviews (n=15), analyzing open-ended data using reflexive thematic analysis and framing findings through activity theory to highlight research's systemic shortcomings. 
 We expose how BLV researchers sacrifice autonomy and shoulder physical burdens, with nearly one-fifth unable to independently perform literature review or evaluate visual outputs, delegating tasks to sighted colleagues or relying on AI-driven retrieval to circumvent fatigue. Researchers also voiced frustration with specialized tools, citing developers' performative responses and losing deserved professional accolades. 
We seek follow-through on research's promises through design recommendations that reconceptualize accessibility as fundamental to successful research and supporting BLV scholars' workflows.
\end{abstract}




\maketitle

\section{Introduction}
\label{sec:introduction}

Researchers represent a minuscule fraction of the global population, approximately 0.12\% according to UNESCO's 2021 Science Report~\cite{unesco_2021,katsnelson_these_2023, who_2023_disability_report, crd_blv_report_2025}~\footnote{The UNESCO Science Report is published every five years, with the next iteration projected to be released in 2026.}. Within this already small community, blind and low-vision (BLV) researchers represent an even smaller fraction, yet they bring unique perspectives to knowledge creation, often developing innovative approaches to research methodology and conceptual frameworks that emerge from their lived experiences navigating research environments not designed with accessibility in mind~\cite{MonaMinkaraBlind, swartzScienceValueDiversity2019, ScienceGoldenInterviews,al-jadir10DisabledScientists2024}. Despite their invaluable contributions, there is limited prior work exploring how BLV researchers successfully craft their practical workflows and complete each step of their research. In other words, how do BLV researchers \textit{do} research? Do BLV researchers use the same tools and workflows as their sighted peers? Are they able to do so in the first place?

Prior investigations identified inaccessibility in specific steps of the research process, including data analysis techniques and collaborative practices~\cite{aishwarya_performing_2022}. Separately, research on academic workflows has highlighted the importance of accessible tools and inclusive practices, especially as the research community becomes increasingly diverse~\cite{hartmann_disability_2019, minkara_implementation_2015, manzoor_assistive_2018, wang_improving_2021}.
However, there remains a limited understanding of how BLV researchers engage in both specific steps of the research process \textit{and} the broader research workflow. While some studies have examined accessibility challenges in research settings~\cite{greenvall_influence_2021, jain_navigating_2020, hofmann_living_2020}, the lived experiences of BLV researchers navigating research tools, collaborative relationships, and institutional barriers have received limited attention. It is crucial to address this gap given documented evidence of declining accessibility in scholarly materials~\cite{kumar_uncovering_2024, singh_chawla_accessibility_2024} and persistent barriers to full participation in research.

Technological advances have transformed research workflows, creating both opportunities and barriers for BLV researchers. The significance of this study lies in its community-driven approach: we, as BLV researchers, aimed to understand how other BLV researchers encounter unique barriers in research participation, an activity fundamental to knowledge creation, which should be accessible to individuals of any ability~\cite{brown_ableism_2018}.
We aim to provide an experiential account of BLV researchers' workflows and processes, including their use of tools and instruments, support systems, workarounds, and collaborative experiences, as well as the challenges they encounter in contributing to their respective disciplines. By examining these experiences through the lens of both individual adaptation and systemic barriers, we strive to present insights that can inform more inclusive research environments.
The following research questions guided our inquiry:

\begin{enumerate}
    \item[\textbf{RQ1)}] What research tools are frequently used among blind and low-vision (BLV) researchers across different stages of the research process (literature review, data collection, analysis, manuscript writing)?

    \item[\textbf{RQ2)}] Which research tools present the most significant accessibility challenges for BLV researchers?

    \item[\textbf{RQ3)}] What strategies do BLV researchers employ to navigate accessibility barriers in their workflows?

    \item[\textbf{RQ4)}] What impacts do research tool accessibility barriers have on BLV researchers' work experiences?

\end{enumerate}

To address these RQs, we conducted an explanatory sequential mixed-methods study with 57 BLV researchers. We first surveyed participants about their tool usage, accessibility experiences, and challenges across the research pipeline, then conducted 15 semi-structured interviews to gain deeper insights into their workflows and strategies. Using activity theory~\cite{activity_theory_2012} as our analytical lens, we examined how BLV researchers navigate systemic barriers in literature review, data analysis, manuscript writing, and dissemination. Our findings reveal that accessibility challenges extend beyond individual tool design to encompass institutional practices, collaborative dynamics, and the structural dependence on commercial software ecosystems that prioritize innovation over basic accessibility.

Our study offers three primary contributions: 

\begin{itemize}
    \item \textbf{Empirical documentation} of accessibility barriers across the research pipeline, including specific tool challenges, workarounds, and their impacts on BLV researchers' productivity and participation.
    
    \item \textbf{Systemic analysis} of why these barriers persist, revealing how commercial software ecosystems, institutional practices, and ableist assumptions create structural inequities rather than isolated technical problems.
    
    \item \textbf{Actionable recommendations} for tool developers, institutions, and research communities to support equitable research participation through accessible research tools, revised evaluation standards, and frameworks that support interdependence.
\end{itemize}

The remainder of this paper proceeds as follows: we first review related work on accessibility in research and disability studies perspectives on research participation (Section~\ref{sec:related_work}), followed by our data collection and analysis methodologies (Section~\ref{sec:methods}). We then present our findings organized around our research questions (Section~\ref{sec:findings}, discuss implications for research workflow design and institutional practices (Section~\ref{sec:discussion}), and conclude with directions for future work in creating more inclusive research environments (Section~\ref{sec:conclusion}).

\section{Related Work}
\label{sec:related_work}


We situate our work within three interconnected bodies of literature. First, we introduce activity theory as our analytical framework, detailing its history and evolution. Second, we examine how systemic barriers in research environments specifically impact BLV researchers, revealing how supposedly neutral research infrastructures function as exclusionary gatekeepers. Finally, we explore how BLV researchers' methodological innovations not only overcome these barriers but also broadly advance research practice. Throughout, we demonstrate how activity theory illuminates these dynamics as \textit{systemic} contradictions rather than \textit{individual} challenges.

\subsection{Theoretical Framework}
\label{subsec:theoretical-framework}


\begin{figure*}[htb!]
    \centering
    \captionsetup{justification=centering}
    \includegraphics[width=0.8\linewidth]{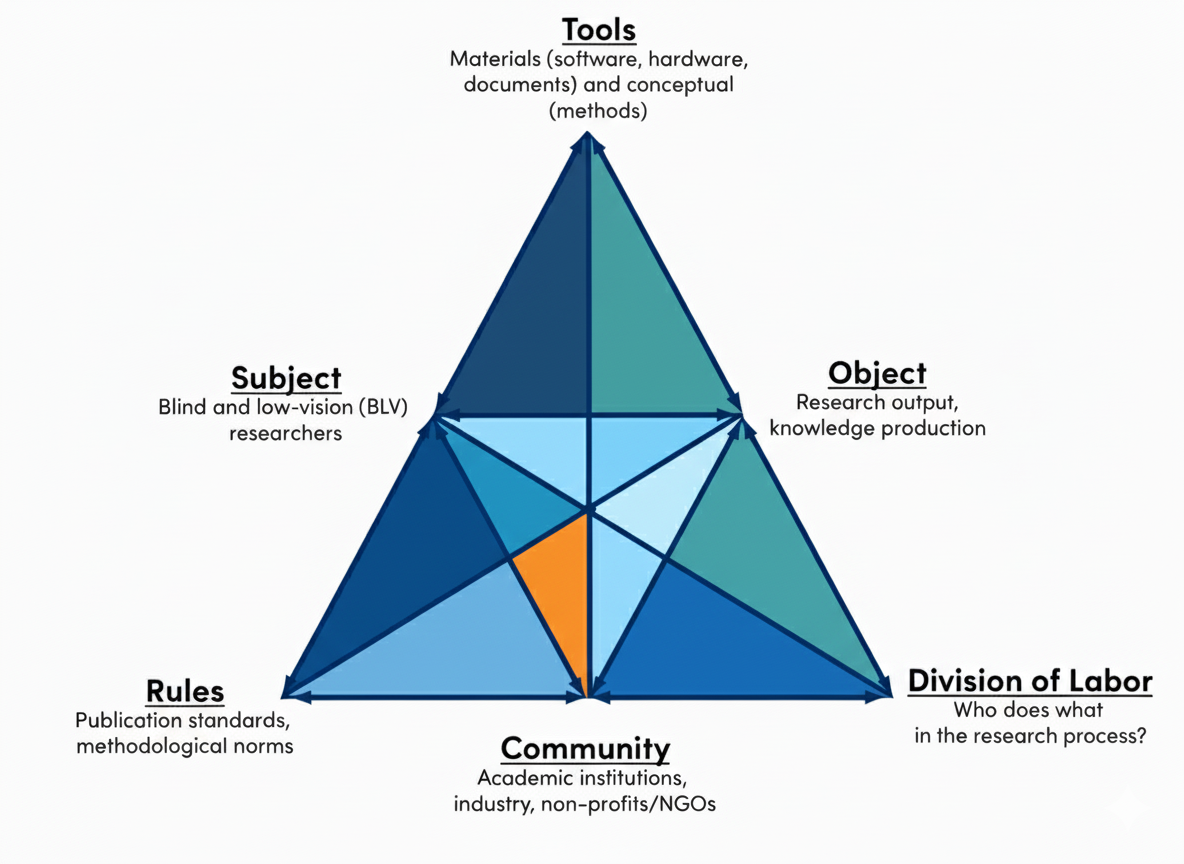}
    \caption{Activity theory core constructs mapped to BLV researchers' workflows.}
    \label{fig:activity-theory-core-constructs}
    \Description{A triangular diagram illustrating the Activity Theory framework for BLV researchers. The triangle has six vertices labeled: Tools (top) - Materials and conceptual methods; Subject (left) - Blind and low-vision researchers; Object (right) - Research output and knowledge production; Rules (bottom left) - Publication standards and methodological norms; Community (bottom center) - Academic institutions, industry, and NGOs; Division of Labor (bottom right) - Who does what in the research process. The center contains interconnected triangular segments in varying shades of blue and teal, with one orange segment.}
\end{figure*}

We begin with an overview of activity theory, its history, and its application to our investigation. While multiple traces of activity theory exist, we start with Lev Vygotsky's work~\cite{vygotsky_1934_thought}. While never using the term \textit{activity theory}, Vygotsky posthumously laid its conceptual groundwork through pioneering \textit{cultural-historical psychology}, emphasizing how human consciousness is shaped through social interaction and mediated by cultural contexts~\cite{vygotsky_1934_thought}. His work emphasized \textit{mediated action}, represented through a triangle connecting subject, tool, and object~\cite{vygotsky_1934_thought, concept_mediation_activity_2021, activity_theory_2012}. He believed that cultural artifacts mediate human-world relationships and that these artifacts evolve throughout history.

Alexei Leontiev expanded on Vygotsky's work to introduce a three-level hierarchy of action: \textit{activities} (oriented toward motives meeting human needs) are composed of \textit{actions} (conscious goal-directed processes that fulfill activities), which are comprised of \textit{operations} (routine, often unconscious processes determined by specific conditions)~\cite{leontiev_1973_problems, hierarchy_human_activity_2021}. This led to \textit{collective activity}, which shows that individual actions serve collective goals through social organization and division of labor, rather than aligning with individual motives~\cite{soviet_psychology_cw, learning_by_expanding}.

This approach supported "third-generation activity theory", which was further refined by Yrjö Engestrom~\cite{engestrom_2014_activity}. His iteration graphically expanded Vygotsky's individual-focused triangle by adding community, rules, and division of labor to the basic subject-tool-object relationship~\cite{engestrom_2014_activity, activity_theory_2012, activity_theory_overview}. This version of activity theory will be the one referenced throughout this paper and is grounded in six core constructs:

\begin{itemize}
  \item The \textbf{tool}: material/non-material artifact that the researcher uses to carry out the action (e.g., qualitative analysis software, screen readers, collaborative writing platforms)
  \item The \textbf{subject}: person carrying out the activity (e.g., blind and low-vision researchers conducting studies)
  \item The \textbf{rules}: norms guiding the activity (e.g., publication standards requiring visual figures, institutional policies about research software)
  \item The \textbf{community}: people involved in the activity (e.g., research collaborators, advisors, peer reviewers, conference attendees)
  \item The \textbf{division of labor}: how the activity is divided between people (e.g., who collects data, who authors manuscripts, who presents findings)
  \item The \textbf{object}: the activity's goal (e.g., producing peer-reviewed publications, advancing knowledge in a field)
\end{itemize}

Activity theory remains relevant for understanding technology-mediated practices. Activity-centric computing, for instance, prioritizes supporting meaningful human activities over organizing systems around technical artifacts~\cite{kaptelinin_2012_activity}. Studies of social media platforms reveal how these technologies function as mediators in collective activities, particularly within specific cultural and communication contexts~\cite{kaptelinin_activity_theory_framework_2018}. In our study, activity theory enables us to analyze BLV researchers' workflows not as individual practices but as activities embedded within, and shaped by, fundamentally inaccessible research ecosystems. This framing reveals how barriers emerge from contradictions between activity system components: when \textit{tools} (research software) fail to support \textit{subjects} (BLV researchers), tensions arise that affect \textit{community} relationships (collaboration with sighted colleagues) and \textit{division of labor} (who performs which research tasks). Similarly, when \textit{rules} (publication standards, institutional policies) assume particular abilities, they create contradictions with the \textit{object} (knowledge production) by excluding certain researchers from full participation. Figure~\ref{fig:activity-theory-core-constructs} maps these constructs to BLV researchers' experiences, showing how such contradictions generate both barriers and innovations. Importantly, this lens shifts focus from individual accommodation to systemic transformation. Barriers are not personal limitations but structural contradictions requiring redesign of the entire activity system.

\subsection{Systemic Barriers in Research Activity Systems}
\label{subsec:access-barriers}


Research, as an activity, has significant heterogeneity across disciplines. Specifically within human-computer interaction (HCI), this has been extensively documented in seminal works~\cite{olson_ways_of_knowing_in_hci_2014}. 
While these works are invaluable for understanding underlying epistemological differences between research methods, we posit that they do not explicitly consider \textit{usability} and \textit{accessibility} of certain research methods, particularly for practitioners with diverse abilities. That is, \textbf{these works embed assumptions about researchers' sensory and physical capabilities}.

The gap between methodological theory and accessible practice becomes apparent when examining the tools researchers rely upon daily. These technologies, often treated as neutral conduits for research activity, instead function as gatekeepers to full participation in knowledge production. For example, Aishwarya reveals how qualitative analysis software forces blind researchers to develop complex, undocumented workarounds~\cite{aishwarya_performing_2022}. Similarly, Manzoor et al. demonstrate how LaTeX's inaccessibility reshapes entire research trajectories, forcing scholars to abandon certain publication venues or collaborative opportunities~\cite{manzoor_assistive_2018}. These individual tool barriers intensify during collaborative research. Das et al. document how collaborative writing platforms force BLV academics and professionals to negotiate tool selection, perform invisible labor to maintain awareness through inaccessible features, and balance advocacy for accessibility against potential social penalties~\cite{das_2019}, while Akter et al. reveal how videoconferencing tools undermine BLV professionals' ability to independently facilitate meetings, requiring reliance on sighted co-hosts and potentially compromising perceptions of professional competency~\cite{akter_2023}. 

Beyond general document formats, mathematical notation presents particularly acute accessibility challenges for BLV STEM researchers. Mathematical content in PDFs, whether rendered as images or through inaccessible LaTeX compilation, remains largely unreadable by screen readers without proper semantic markup like MathML~\cite{karshmer_2002, suzuki_2003, hayes_2024}. Moreover, these digital accessibility barriers intersect with physical infrastructure challenges. Transportation limitations and inaccessible fieldwork sites create compounding exclusions~\cite{chiarella_2020, stokes_2019}, particularly affecting researchers with multiple marginalized identities~\cite{goethals_2015, brinkman_2023}. These insurmountable and intersecting barriers function not as discrete obstacles but as interlocking systems of exclusion, where each barrier amplifies others to create multiplicative rather than additive effects to participation in knowledge production.

Outside of individual tools, research environments embody what Brown et al. term "ableist infrastructure", systems built on normative assumptions about researcher capabilities that create contradictions across all elements of research~\citet{brown_ableism_2018}. This manifests across multiple dimensions: inaccessible conference formats~\cite{rizzo_accessible_2024}, communication technologies that fail disabled users~\cite{singh_chawla_accessibility_2024,bigham_uninteresting_2016}, and physical spaces that exclude disabled researchers~\cite{katsnelson_these_2023}. These infrastructural barriers then compound into \textit{epistemic violence}, the systematic dismissal of disabled researchers' knowledge claims~\cite{dotson_tracking_2011,scully_she_2018}. As Ymous et al. document, disabled HCI researchers face "terrifying" futures when their embodied expertise conflicts with established norms, leading to internalized ableism where researchers blame themselves for systemic failures~\cite{ymous_i_2020, campbell_internalised_2009}.

Understanding these barriers through activity theory~\cite{engestrom_2014_activity, activity_theory_2012, activity_theory_overview} reveals them as fundamental contradictions within research activity systems. For instance, inaccessible collaborative tools create tensions between the \textit{division of labor} (who can lead meetings or co-author) and the \textit{community} (collaborative relationships), forcing BLV researchers to either exclude themselves from certain roles or perform additional invisible work. Similarly, inaccessible conference formats represent contradictions between \textit{rules} (presentation norms) and the \textit{object} (knowledge dissemination), where assumed capabilities embedded in presentation standards prevent certain researchers from sharing their work. This systemic framing highlights that accessibility is not about individual researchers' limitations but about how research activity systems are structured to exclude.

Importantly, these innovations align with cripistemological frameworks~\cite{mcruer_proliferating_2014} that position "disabled ways of knowing" as essential to comprehensive understanding. Wang et al. empirically demonstrate this principle: accessibility improvements in scientific documents benefit \textit{all} researchers through enhanced navigation and content extraction. This finding suggests that inclusive design strengthens rather than compromises research quality, with profound implications for research infrastructure. While these barriers create significant challenges, BLV researchers do not passively accept exclusion. Instead, as we explore next, they actively develop innovations that not only enable their participation but often enhance research practice for all.

\subsection{Towards Re-Imagining Equitable Research Participation}
\label{subsec:participating-in-research}


Contrary to deficit narratives, disabled researchers consistently advance methodological rigor through innovation, transforming contradictions within activity systems into opportunities for systemic improvement. Jain et al.'s travel autoethnography reveals insights unavailable through traditional methods~\cite{jain_autoethnography_2019}, while their work on accommodation failures demonstrates how inaccessibility generates alternative strategies that enhance methodological flexibility~\cite{jain_navigating_2020}. Hofmann et al. similarly document how disabled researchers' innovations surpass conventional approaches in both rigor and insight~\cite{hofmann_living_2020,castrodale_youre_2015}. 

These innovations reflect a fundamental shift in recent scholarship from accommodation models toward frameworks of interdependence and participatory design, which recognize that all researchers rely on tools, collaborators, and infrastructures, making accessibility a universal rather than special concern~\cite{bennett_interdependence_2018, van_der_velden_participatory_2021}. This re-framing transforms accessibility from individual subject modification to systemic transformation of the entire activity system. 

In the case of BLV researchers, Sahtout argues that supporting visually impaired researchers requires systematic infrastructure changes~\cite{sahtout_how_2020}. Emerging initiatives demonstrate practical pathways, such as accessible lab design that re-imagines physical tools and spaces~\cite{trager_visionary_2024}, improved PDF remediation that addresses infrastructural inaccessibility~\cite{schmitt-koopmann_more_2025}, and inclusive conference protocols that restructure community norms of academic exchange~\cite{katsnelson_these_2023}. 

However, these remain fragmented and reactive. There is a critical lack of empirical grounding in BLV researchers' actual practices. We address this gap through detailed evidence of BLV researchers' practices, revealing specific contradictions within workflows, and identifying concrete design opportunities for systemic change. Through activity theory, we analyze these tensions not as individual failures but as contradictions demanding systemic transformation. When BLV researchers develop workarounds, such as negotiating alternative collaboration tools or redesigning data collection methods, they reveal contradictions between current research \textit{tools} and the needs of diverse \textit{subjects}, pointing toward necessary infrastructure changes. When they innovate methodologically, they transform contradictions within the activity system into opportunities for improvement, demonstrating that accessibility innovations strengthen research practices for all. Activity theory thus provides both a diagnostic lens for identifying systemic barriers and a generative framework for envisioning more inclusive research ecosystems.
\section{Methods}
\label{sec:methods}

We conducted an explanatory sequential mixed-methods study to investigate how BLV researchers navigate and transform research activity systems. This design enabled us to capture breadth through surveys (n=57), then develop depth through semi-structured interviews (n=15). Activity theory guided both our methodological approach and analytical framework, allowing us to examine research as a collective activity system rather than isolated individual practices.

\subsection{Positionality and Reflexivity}
\label{subsec:positionality}

As BLV researchers ourselves, we bring both methodological strengths and potential biases to this work. The first author has low vision, and the second author is totally blind, providing experiential diversity that enhanced our ability to recognize varied navigation strategies and challenge assumptions about visual access. This positionality functioned as a methodological asset, enabling rapport with participants and sensitivity to subtle accessibility challenges that sighted researchers might overlook~\cite{mankoffDisabilityStudiesSource2010, sharifShouldSayDisabled2022}.

\subsection{Participant Recruitment}
\label{subsec:participants}

We recruited BLV researchers across disciplines, career stages, and geographic regions through multiple channels. To have any chance at capturing such a participant pool, we partnered with several BLV advocacy organizations across the world, including U.S.-based organizations, including the National Federation of the Blind~\footnote{https://www.nfb.org}, the American Foundation for the Blind~\footnote{https://www.afb.org/}, the American Council of the Blind~\footnote{https://www.acb.org/home}, and the University of Washington's DO-IT Center~\footnote{https://doit.uw.edu/}. Outside of the United States, we contacted the Blind Academics mailing list, the International Council of Education for People with Visual Impairments (ICEVI)~\footnote{https://icevi.org/}, and the World Blind Union (WBU)~\footnote{https://worldblindunion.org/}. We first reached out to these organizations for approval to advertise our study to their networks, as well as snowball sampling~\cite{goodman_snowball_1961} by inquiring about other community partners. Once approved, either the first author sent out recruitment information to prospective participants via email, or an organization representative sent it on our behalf. Participants had to meet the following inclusion criteria to participate in both phases of the study:

\begin{itemize}
    \item Participants \textbf{must be 18 years of age or older}.

    \item Participants \textbf{must identify as either being legally blind or having low vision (visual acuity of 20/70 or less)}.

    \item Participants \textbf{must be actively engaged in research} \textit{or} \textbf{have previously engaged in research within any academic discipline}, \textbf{and within any context} (e.g., academia, industry, independent researcher, etc.). 
\end{itemize}

\subsection{Survey Design}
\label{subsec:survey_design}

We structured our survey based on the stages of the research process (literature review, data collection, data analysis, writing, and dissemination) as proposed by Creswell and Creswell~\cite{creswell_research_2017} to capture BLV researchers' experiences throughout their processes. We synthesized three theoretical frameworks when designing our survey questions to comprehensively capture BLV researchers' experiences while also answering our RQs: the Technology Acceptance Model (TAM)~\cite{davis_technology_1989}, the User Experience Questionnaire (UEQ)~\cite{laugwitz_construction_2008, laugwitz_subjektive_2009}, and activity theory~\cite{engestrom_2014_activity}. We chose these frameworks for their abilities to measure technical and non-technical aspects of user satisfaction and to provide insights into the societal contexts in which they leverage particular technologies. Specifically, we drew inspiration from the TAM's constructs of perceived usefulness (PU) and perceived ease of use (PEOU) constructs to examine both technical tools and social systems~\cite{davis_technology_1989}. Questions assessed not only software accessibility but also how institutional policies and collaborative practices influence research participation. We extended the UEQ's six dimensions (attractiveness, perspicuity, efficiency, dependability, stimulation, novelty) beyond technical interfaces to encompass the broader research ecosystem. This allowed examination of how research environments stimulate or constrain BLV researchers' innovation~\cite{laugwitz_construction_2008, laugwitz_subjektive_2009}. 

Foregrounding our survey's theoretical approach to understanding BLV researchers' workflows was activity theory~\cite{engestrom_2014_activity}, which informed question design based on its six core constructs. We chose this theory as our study's guiding analytical lens for several reasons; first, research is an inherently dynamic activity, involving diverse stakeholders with whom the researcher must engage to achieve a set of desired goals. The pathways through which the researcher can effectively engage with collaborators to make their contributions may seem infinite, but this is seldom the case, particularly for BLV practitioners who must leverage a finite set of options to participate in collaborative work~\cite{sahtout_how_2020}. Second, activity theory pays special attention to \textit{intentional} and \textit{purposeful} activity, which is essential to decision-making across all stages of the research process~\cite{engestrom_2014_activity, creswell_research_2017}. Such thought processes of BLV researchers have received limited attention~\cite{aishwarya_performing_2022}, and our RQs sought to address this gap. Third, our interest in the types of research \textit{tools} used by BLV researchers from both technical and socio-technical perspectives is given particular attention in activity theory. Specifically, activity theory examines how tools (in a technical sense) enable the subject(s) to achieve their object(s) within a particular community that inevitably has its own set of established norms. Appendix~\ref{sec:survey-instrument} provides a detailed breakdown of our survey design and its usage of the above frameworks.

\subsection{Interview Design}
\label{subsec:interview_design}

Our interview design was grounded in our desire for BLV researchers to have an open, safe space to share their unique, lived experiences with research. To balance structure and openness of our conversations, we chose semi-structured interviews (refer to Appendix~\ref{sec:interview-script}) for their ability to provide a flexible roadmap for discussion, ensuring that participants could discuss aspects of the research process unique to them while simultaneously gathering pertinent insights for our RQs.    

\subsection{Data Collection}
\label{subsec:data-collection}

In the first phase of the study, participants completed an online survey (refer to Section~\ref{sec:survey-instrument}) via Qualtrics~\cite{qualtrics}, gathering high-level details about individual workflows throughout the research cycle (RQ1, RQ2). At the survey's end, participants could opt into a paid follow-up interview with a research team member. Survey data collection took place from June 2025 to July 2025. 

The first author then contacted participants interested in the second phase of the study via email. Participants chose a one-hour time block for the interview, and they received an email confirmation with the informed consent documents, the scheduled time, date, and online meeting link. Before each interview, the first author obtained verbal confirmation and consent from each participant, verifying they had reviewed the study procedures and agreed to the session being audio-recorded. The first author remotely conducted all interviews from July 2025 to August 2025. All interview participants indicated they were actively engaged in research at the time of their interviews, with research experience ranging from 2 to 40+ years. Most participants ($\tfrac{11}{15}$; 73\%) began their research careers after 2010, ensuring our findings predominantly reflect accessibility challenges within modern research ecosystems rather than outdated technologies. However, three senior participants with 20+ years of experience provided valuable longitudinal perspectives on how barriers have persisted despite technological advances~\footnote{P9 declined to give an estimate for years of research experience, but indicated "decades" of prior experience in academic research and present-day involvement in legal research.}. Participants were compensated with a USD 30 Amazon e-gift card approximately one week after their interview.

This study received IRB approval from the University of Illinois. Given the professional risks of discussing workplace accessibility barriers, we implemented robust confidentiality protections. All participants were assigned pseudonyms, and we removed institution names, specific job titles, and other identifying information from transcripts. Participants were explicitly informed during consent that they could skip questions, decline to name institutions, or withdraw at any time. Audio recordings were stored on university-owned, password-protected computers and deleted after transcription. Only the research team had access to identifiable data, which was stored separately from the de-identified transcripts. We emphasized participant agency throughout, with several participants declining to discuss specific incidents when disclosure could compromise their professional standing.

\subsection{Data Analysis}
\label{subsec:data-analysis}

Our explanatory sequential design employed a two-phase analytical approach where quantitative findings from the survey informed and focused our subsequent qualitative investigation. In phase 1, we analyzed closed-ended survey responses using descriptive statistics (measures of central tendency and variability) and non-parametric tests (Kruskal-Wallis~\cite{kruskal_use_1952}) to identify research stages with the greatest perceived difficulty and map the landscape of tools used across each research stage, as defined in Section~\ref{subsec:survey_design}. These quantitative patterns guided our interview protocol design, enabling us to probe specific high-barrier activities and underexplored tool categories in phase 2. 

For phase 2, we employed thematic analysis that integrated deductive and inductive elements~\cite{fereday_demonstrating_2006}. All transcripts were transcribed using OpenAI's Whisper~\cite{whisper}, with participants informed and provided verbal consent before each interview. The first author conducted initial open coding to identify emergent concepts, then iteratively organized codes into themes that both addressed our research questions and captured unexpected patterns from participant narratives. This dual approach, using RQs as an organizing framework while remaining open to unanticipated findings, enabled us to systematically explore predefined areas of interest (tool usage, barriers, strategies, impacts) while incorporating participant experiences that extended beyond our initial theoretical scope. All authors collaboratively refined themes through discussion until consensus was reached, resulting in four final themes.

We also conducted qualitative analysis of open-ended survey responses using systematic coding in ATLAS.ti 25~\cite{ATLASTI_2025}. We followed a three-stage process: initial open coding to identify emergent concepts, axial coding to establish relationships between concepts, and selective coding to align responses with activity theory's six core tenets~\cite{engestrom_2014_activity}. These open-ended responses provided initial context for barriers identified quantitatively and helped refine our interview questions.

Our integration occurred at two points: 1) quantitative results identifying high-barrier stages shaped which aspects we explored deeply in interviews, and 2) qualitative findings from both surveys and interviews explained \textit{why} particular stages or tools emerged as problematic in quantitative analyses. This sequential integration enabled us to transition from identifying existing barriers (phase 1) to understanding how and why they manifest in the workflows of BLV researchers (phase 2).

\section{Findings}
\label{sec:findings}

We now present our findings from 57 survey respondents (refer to Appendix~\ref{subsec:survey-participants} for demographics) and 15 interviews (refer to Appendix~\ref{subsec:interview-participants} for demographics), organized by our RQs, shaped by reflexive thematic analysis~\cite{braun_using_2006}, and grounded in activity theory~\cite{engestrom_2014_activity}. We show how accessibility barriers at each research stage, often embedded in tools themselves, cascade through BLV researchers' workflows, excluding them from knowledge production. 

\subsection{Research Tool Choices Among BLV Researchers (RQ1)}
\label{subsec:rq1-findings}

\begin{figure*}[t]
    \centering
    \captionsetup{justification=centering}
    \includegraphics[width=0.6\linewidth]{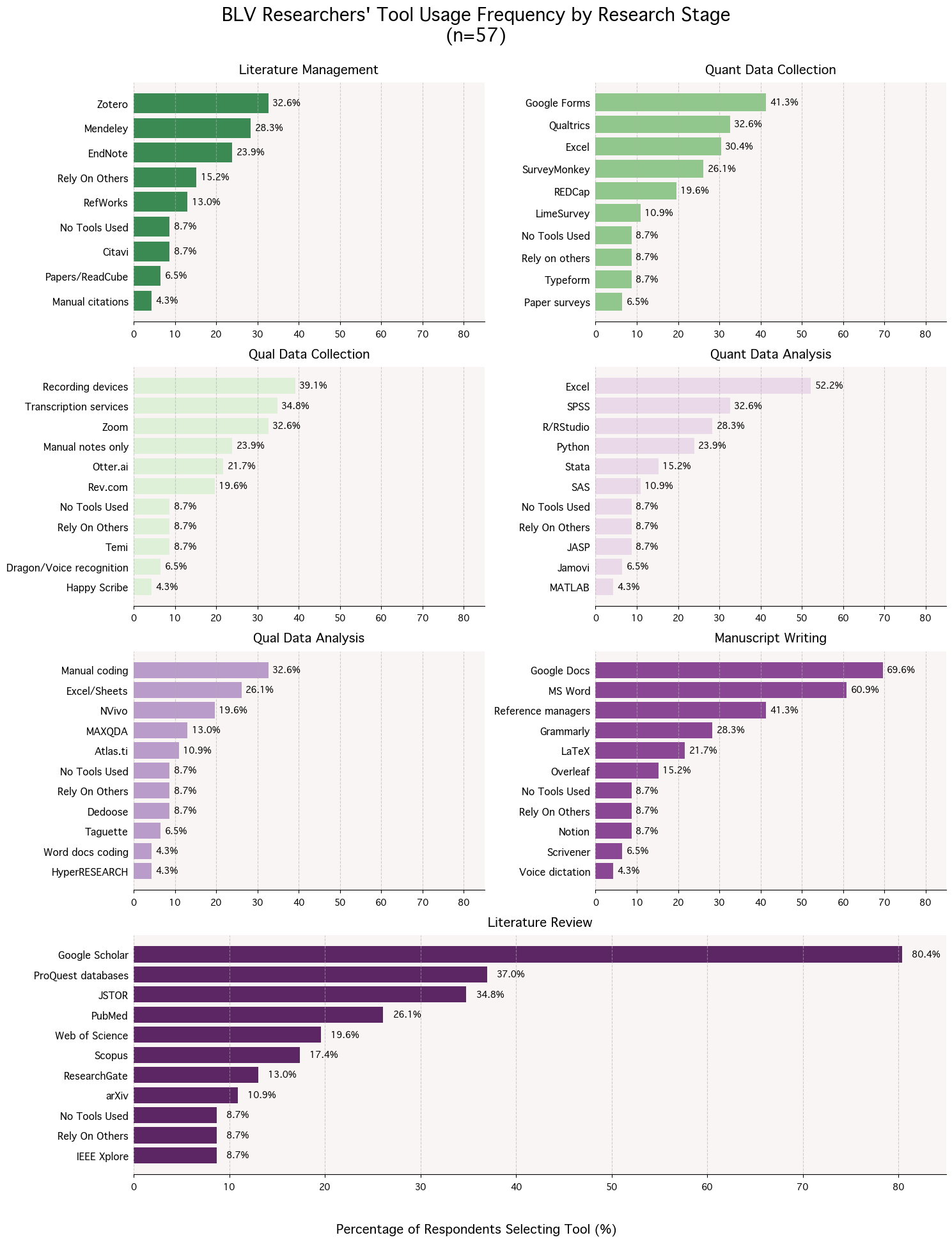}
    \caption{Distribution of tool usage among participants (n=57).}
    \label{fig:tool_distribution}
    \Description{Seven horizontal bar charts showing tool usage frequency by BLV researchers (n=57) across research stages. Literature Management: Zotero (32.6\%), Mendeley (28.3\%), EndNote (23.9\%). Qualitative Data Collection: Recording devices (39.1\%), Transcription services (34.8\%), Zoom (32.6\%). Quantitative Data Collection: Google Forms (41.3\%), Qualtrics (32.6\%), Excel (30.4\%). Qualitative Data Analysis: Manual coding (32.6\%), Excel/Sheets (26.1\%), NVivo (19.6\%). Quantitative Data Analysis: Excel (52.2\%), SPSS (32.6\%), R/RStudio (28.3\%). Manuscript Writing: Google Docs (69.6\%), MS Word (60.9\%), Reference managers (41.3\%). Literature Review: Google Scholar (80.4\%), ProQuest databases (37.0\%), JSTOR (34.8\%).}
\end{figure*}

When investigating RQ1, we relied on participants' survey responses for which tool(s) they use during each research stage (refer to Figure~\ref{fig:tool_distribution}). BLV researchers consistently choose general-purpose tools (Word, Excel, Google Docs) over specialized research software (tools for specific research purposes e.g., NVivo, ATLAS.ti, etc.). Particularly in the case of literature review and management, we found that nearly one-fifth of participants (17\%) reported either delegating these tasks to sighted colleagues or using AI-driven retrieval to circumvent fatigue. 

As we later uncovered through follow-up interviews, stage-specific frustrations were common. For literature review, participants described usability and learnability challenges with \textit{reference management tools}, particularly with users' existing ATs. As P3, a doctoral candidate in speech education, noted: 

\begin{quote}
    "\textit{If I was taught more explicitly how to use [them] with JAWS...I might be using [them] to [their] full functionality.}" - P3
\end{quote}

P12's accounts reinforced this sentiment, who preferred to track citations using a single online document: 

\begin{quote}
    "\textit{I think I fall back to probably the more tedious way because I know it works.}" - P12
\end{quote}

Another challenging aspect of literature review was \textit{PDF accessibility}, as P2 described: 

\begin{quote}
    "\textit{The problem with these materials can be that the OCR is not done well, that those scanned publications can sometimes be pretty illegible or hard to read.}" - P2
\end{quote}

P3 elaborated on these challenges through their access experiences with older literature:

\begin{quote}
    "\textit{the accessibility of literature itself also really varies depending on when it was published how it was released.}" - P3
\end{quote}

For data analysis and visualization, multiple participants independently developed workarounds, reflecting pragmatic adaptation rather than preference. P4, a research director, described this convergence:

\begin{quote}
"\textit{I think the most creative workaround is probably, although I guess it's not original because I know other [blind] people do the same thing, using Word or Excel for qualitative coding instead of proprietary tools like NVivo or ATLAS.ti.}" - P4
\end{quote}

This trust erosion in specialized tooling was a recurring theme; P1, an economics doctoral candidate, articulated the underlying dynamic that drove them towards general-purpose tools:

\begin{quote}
    "\textit{\textbf{I think I don't trust any technology or any professional research tool at all}...once I know how to use it, even just one hour, maybe two weeks later, they got an update [and] immediately my screen reader cannot read out something properly.}" - P1
\end{quote}

The constant threat of accessibility regression forces researchers to prioritize stability over functionality, fundamentally constraining their methodological choices. Activity theory reveals this as a systemic contradiction: inaccessible tools create tensions not only between tool and subject (researchers unable to execute tasks) but also between subject and object (researchers unable to achieve knowledge production goals), between tools and division of labor (workarounds requiring sighted colleague assistance), and between rules and community (institutional software mandates excluding BLV researchers from collaborative workflows). This cascade of contradictions transforms what appears as individual accommodation into structural exclusion -- the activity system itself, not individual capability, determines who can participate in research.

\subsection{Critical Access Barriers Manifesting Across Research Stages (RQ2)}
\label{subsec:rq2-findings}

\begin{figure*}[t]
    \centering
    \captionsetup{justification=centering}
    \includegraphics[width=0.7\linewidth]{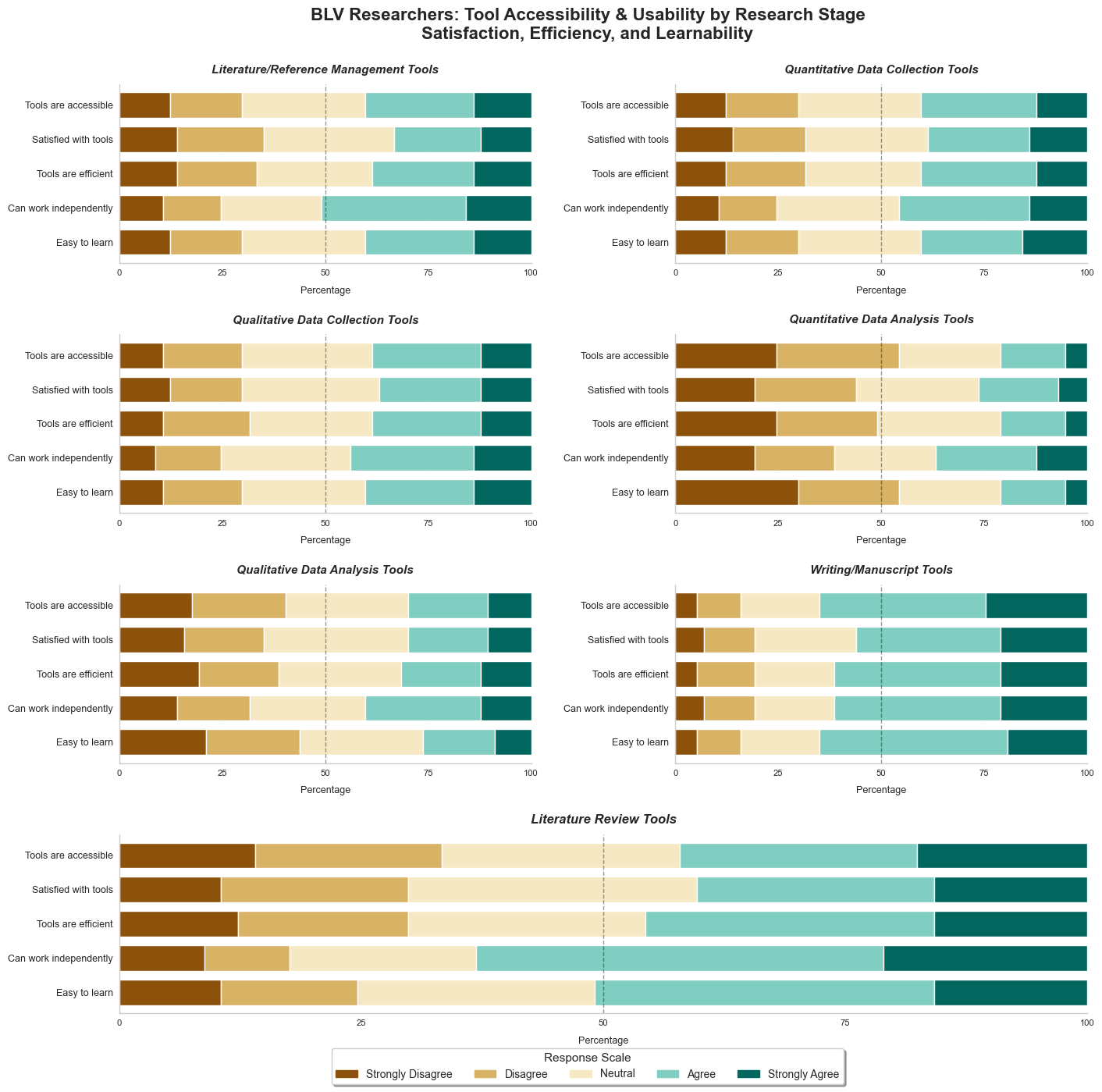}
    \caption{Distribution of tool Likert scale ratings among participants (n=57).}
    \label{fig:tool_likert_scale_distro}
    \Description{Seven horizontal diverging bar charts showing satisfaction ratings for research tool categories used by BLV researchers. Each chart displays five satisfaction metrics (Tools are accessible, Satisfied with tools, Tools are efficient, Can work independently, Easy to learn) across seven tool categories (Literature/Reference Management, Quantitative Data Collection, Qualitative Data Collection, Quantitative Data Analysis, Qualitative Data Analysis, Writing/Manuscript Tools, Literature Review Tools). Responses use a 5-point scale from Strongly Disagree to Strongly Agree, with most categories showing neutral to positive ratings and Literature Review Tools receiving the highest positive ratings.}
\end{figure*}

\begin{figure*}[h]
    \centering
    \captionsetup{justification=centering}
    \includegraphics[width=0.8\linewidth]{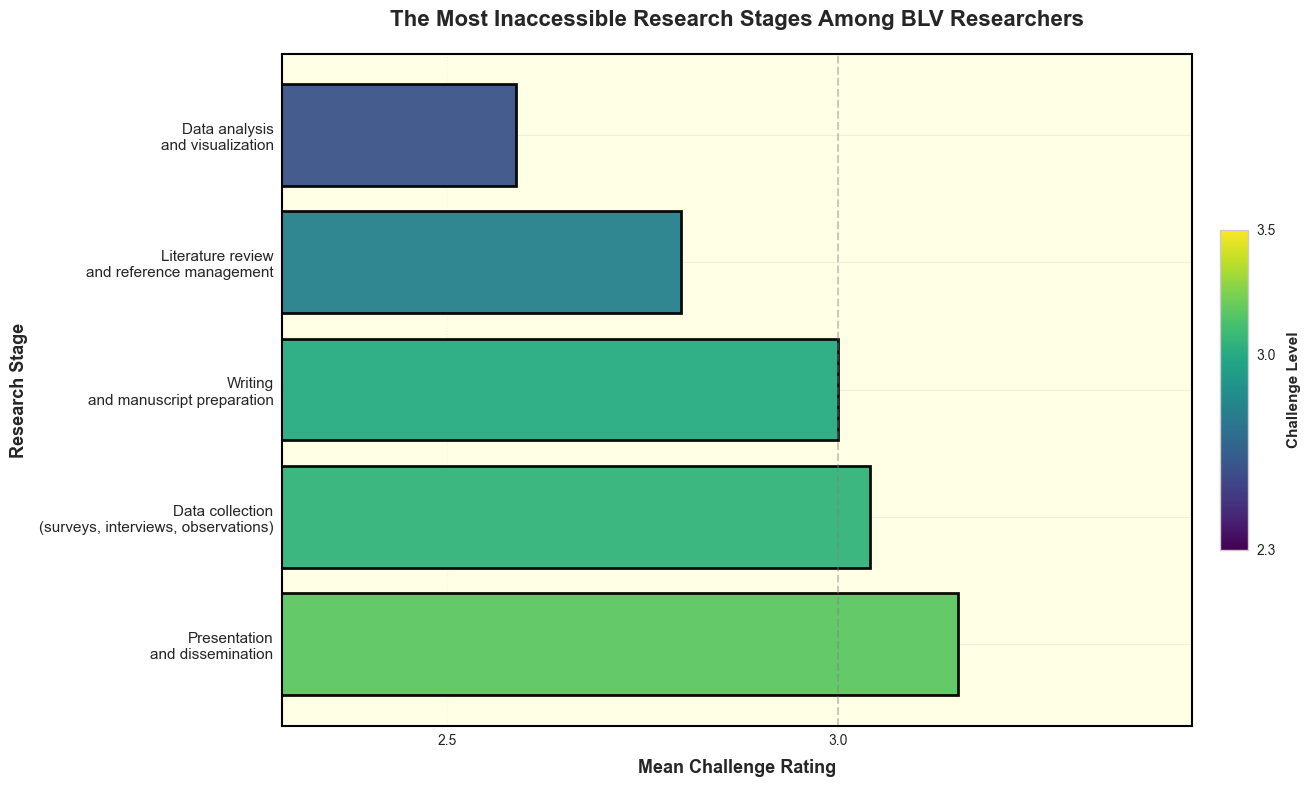}
    \caption{Mean accessibility challenge ratings among participants, from 1 being most inaccessible to 5 being most accessible (n=57).}
    \label{fig:q37_inaccessible_research}
    \Description{Horizontal bar chart displaying the most inaccessible research stages for BLV researchers, ranked by mean challenge rating. From most to least challenging: Presentation and dissemination (3.3), Data collection including surveys, interviews, and observations (3.0), Writing and manuscript preparation (2.9), Literature review and reference management (2.7), and Data analysis and visualization (2.4). Challenge levels are indicated by color gradient from dark purple (2.3) to yellow-green (3.5), with higher values indicating greater inaccessibility.}
\end{figure*}

To address RQ2, we asked participants a series of 5-point Likert scale questions to assess their experience with their selected tools for each research stage (refer to Figure~\ref{fig:tool_likert_scale_distro}) and ranking the stages from 1 (most challenging) to 5 (least challenging). Figure~\ref{fig:tool_likert_scale_distro} depicts the rating distribution; data analysis and visualization ranked as most challenging ($\bar{x}$ = 2.58/5), followed by literature access ($\bar{x}$ = 2.80/5). P12, a music instructor, explained this challenge that no traditional assistive technology addresses:

\begin{quote}
    "\textit{The barrier is \textbf{how do I know if it looks good visually?}...I can enter it all in there, but that doesn't mean it's all going to look good for someone else who's trying to look through data.}" - P12
\end{quote}

Additionally, when leveraging specialized research tools, participants noted consistent, basic accessibility failures that blocked them from performing this critical research step. P3 shared their  experiences with qualitative analysis software:

\begin{quote}
    "\textit{I was taking more qualitative data analysis classes. And we were taught NVivo. And very quickly realized like NVivo was not really accessible...I emailed the company and told them like, oh, I cannot use your program, what are some workarounds. And they actually emailed me to say, oh, have you tried these tricks...have you tried these things. And I was like, yes, I have tried these things and it actually doesn't work. Once I open your program, I cannot even close it. And...they then responded like, we'll see what we can do...But yeah, they also emailed me back about a year later like, oh, we couldn't figure anything out at this time. So we're going to close the case.}" - P3
\end{quote}

However, P3's frustration extends beyond individual software to the broader research tool landscape:

\begin{quote}
    "\textit{We are making self driving cars right...And I think figuring out how to analyze qualitative data, which is essentially a string of text...for that to be not accessible just feels absurd. \textbf{If we can make cars that drive without human input...analyzing text data should be a no-brainer}...For me to not have the options that other people have seems very wrong.}" - P3
\end{quote}

Through activity theory, we see these events as obstacles across multiple research stages. Inaccessible tools prevent BLV researchers from completing the objects of their research activities, which in turn have ripple effects throughout the rest of the activity system. P3's experience with NVivo highlights how the tooling itself imposes constraints on the researchers' ability to perform their research, \textit{not} the individual researcher's ability. These tools may have been created without considering BLV researchers' access needs, and are thus unable to empower a more diverse research community.

Follow-up interviews also revealed the inaccessibility of manuscript submission and review platforms, particularly for P8:

\begin{quote}
    "\textit{On the side of submitting for publication and then reviewing, the platforms that we use are really not that accessible...I've had to contact a journal editor. I was reviewing a paper and I couldn't, for some reason, the way the tabs opened and stuff like I couldn't read the revised paper or the responses that the authors put. And so he had to email me the document separately.}" - P8
\end{quote}

P8's experience captures how tool inaccessibility creates systemic contradictions: platforms unusable by BLV reviewers (tool-subject) necessitate editor intervention, shifting division of labor from independent to mediated review while conflicting with community expectations of autonomous workflows. These cascading contradictions prevent capable researchers from achieving research objects (peer review, knowledge dissemination) not through individual limitation but through structural exclusion embedded in the activity system itself.

\subsection{Adaptive Strategies Reveal Hidden Labor (RQ3)}
\label{subsec:rq3-findings}

To investigate RQ3, we asked participants about their creative strategies to overcome accessibility barriers. We found that BLV researchers develop workarounds that expose both resilience and problematic dependencies. One frequently used workaround was \textit{AI, specifically large language models (LLMs)}. Participants shared how emerging AI tools offer unprecedented workarounds for previously insurmountable barriers. P11, a gender studies graduate student, described their LLM usage with inaccessible PDFs:

\begin{quote}
    "\textit{I just downloaded and then uploaded to ChatGPT and then they can just tell you page by page...'Can you tell me what is the first page of this document and give me the description?'}" - P11
\end{quote}

Though requiring multiple queries for tasks sighted researchers accomplish at a glance and raising pre-publication confidentiality concerns, this represents progress over previous options. However, it also underscores the access barriers that BLV researchers experience when using resources that their non-BLV peers may take for granted. BLV researchers employ these alternatives to work towards the objects of their research, highlighting their perseverance.

Separately, another recurring strategy consisted of \textit{forming trust networks and contingency planning}. The precarity of accessibility across research tooling has the potential to generate profound psychological burdens. P1 described their constant vigilance:

\begin{quote}
    "\textit{I keep checking what other friends are using and make sure I know what's the backup plan...Who knows two months later what's going to happen?}" - P1
\end{quote}

This vigilance represents significant hidden labor, manifesting not just in their primary workflows but in multiple contingency systems. BLV researchers must switch to a different tool or strategy at a moment's notice, as the tools they use may suddenly become inaccessible.

Researchers also shared the benefits and tradeoffs of \textit{collaborative compromises}. Task distribution often reflects accessibility constraints rather than expertise, even when BLV researchers have the expertise. P14, a plasma physicist, offered a stark assessment:

\begin{quote}
    "\textit{The easiest [workaround] is to hire younger people to do the work...as they set up the instrumentation and collect the data}" - P14
\end{quote}

While P14 is likely just as capable (if not more so) to accomplish these tasks, delegating portions of their responsibility to younger colleagues is a pragmatic solution to the accessibility barriers they face. It may help them in the short term, but it also limits their ability to sharpen skills they need to hold themselves in high regard, particularly when seeking promotion. P11 confirmed how accessibility shapes collaboration patterns, even though they may be interested in other tasks:

\begin{quote}
    "\textit{I think in part it may be because of the accessibility challenges that I'm the one who did more of the interviews...I was not the one designing the posters, even though I wanted to help.}" - P11
\end{quote}

P11's specialization maintains team productivity but limits their opportunities to develop diverse skills expected for advancement. While they may prefer their assigned or chosen tasks in some cases, it nonetheless raises questions about whether there are more equitable approaches to division of labor. BLV researchers should have equal opportunities to pursue tasks of their choosing, rather than being limited to a predefined set.

\subsection{Compound Professional Impacts (RQ4)}
\label{subsec:rq4-findings}

For RQ4, we examined how systemic accessibility barriers throughout the research community reshape entire careers. One alarming, persistent phenomenon was \textit{professional network limitation and degradation}, or the inability to expand professional networks. P14 talked about their first-hand experience of trying to find speakers after a seminar:

\begin{quote}
    "\textit{If I wanted to go talk to the speaker after a seminar and ask some offline questions, that's very difficult...particularly if it's in say a conference setting where there might be, you know, a dozen speakers in a session...and then I have to go, you know, 50 people in the room, and I've got to go find that person.}" - P14
\end{quote}

Such micro-barriers accumulate into macro-level professional isolation. The research community fails BLV researchers who cannot access its resources. BLV researchers also face \textit{credibility challenges} that undermine their standing in research. That is, participants frequently reported how others may make immediate assumptions about their ability to contribute. For example, P9, a lawyer, shared their frustrations with having to re-prove their qualifications to non-BLV colleagues:

\begin{quote}
"\textit{The major impediment for professional networking is the attitude, not the accessibility. [Some people] keep summarizing. As in, they have their own preconceived ideas about somebody being blind, despite them acting differently in the public sphere....\textbf{They might not feel or believe that you can contribute.} We should not have to labor or make an effort twice or thrice as compared to an ordinary person in the same situation. Just to convince somebody that doesn't even have a better qualification than yourself.}" - P9
\end{quote}

These quick judgments reflect broader ableist assumptions~\cite{marathe_2025} about BLV researchers' capacity for meaningful research contributions. It may also be indicative of a systemic lack of understanding of the labor and expertise required to produce successful research, assuming that certain physical abilities are required to contribute to knowledge production, creating barriers to equal participation in research.

Separately, participants expressed frustrations over \textit{loss of professional accolades}. P1's conference experience exemplifies this challenge:

\begin{quote}
    "\textit{Data visualization is really out of my control...I try to make my diagrams visually appealing...but still I have no clue what beautiful means [and] what is easy to understand. I try to ask AI...[about] what is a visually appealing diagram, but since I'm not able to see it clearly myself, I am still not able to fully digest and use....There was a conference and they went to select the best presentation award. I don't think my presentation was worse than anyone else in the same conference but I [didn't] use very visually appealing pictures in it. Just because of my visual impairment, I don't have the opportunity to do something like others [and] I will feel a little bit unhappy about it.}" - P1
\end{quote}

In this case, P1's frustration goes beyond the tools themselves and is instead directed at implicit expectations surrounding successful research. P1's presentation may have been of the same (if not better) caliber than their peers, but due to uncontrollable physical factors, they lost out on prestigious accolades that will eventually differentiate them from other candidates on the job market. This account in particular highlights how evaluation metrics are inherently ableist, as they fail to account for the hidden labor required to produce successful research, especially for BLV researchers. Activity theory captures this as tension between subjects (BLV researchers) and rules (evaluation metrics), which may potentially create community-level inequalities.

However, most alarmingly, BLV researchers may experience \textit{exclusion from research-oriented careers altogether} due in part to the uncertain long-term nature of their vision. P7, a public health researcher, shared their intricately crafted career trajectory:

\begin{quote}
    "\textit{Even getting into public health, I have gone on the route that I have because I've been pretty confident that I as my vision loss progresses and as I am now...[I ask myself], can you, you know, keep like working at a good capacity...but now I'm like, well, maybe I can't do that anymore}" - P7
\end{quote}

Moreover, P13, a non-profit CEO, shared their journey and how their experiences epitomize their vision's social costs:

\begin{quote}
"\textit{My sphere of influence and networking has been greatly reduced since I've developed my disability over the last 10 years. I had to step back from the field that I was going to go into and all of the influence that I had within that field, you know, it's gone now. \textbf{I had two classes and a dissertation shy of completing [my PhD]...so close and yet I had to walk away.}}" - P13
\end{quote}

P13's story exemplifies the ultimate cost of these accumulated barriers -- inaccessibility influenced their decision to abandon a research career altogether. Through activity theory, these findings reveal not individual failures but systemic contradictions. The tool-subject contradiction drives adoption of suboptimal alternatives. The rules-community contradiction (evaluation metrics ignoring accessibility labor) creates invisible disadvantages. The division of labor contradiction (accessibility-driven task allocation) limits professional development. These contradictions identify key areas where inclusive design, policy changes, and systemic reforms can enable equitable participation in knowledge production.
\section{Discussion}
\label{sec:discussion}

Our study reveals how the research activity system creates cascading access barriers for BLV researchers, fundamentally altering their workflows. Through activity theory~\cite{engestrom_2014_activity}, we identify systemic contradictions extending beyond individual tool failures to challenge how we, as a community, can improve research accessibility. Our conception of "research" extends beyond desk-based activities, reading papers, writing manuscripts, interacting with software, to encompass the full ecology of practice: laboratory equipment, field sites, conference navigation, and transportation. Each dimension presents distinct accessibility challenges that compound to shape BLV researchers' career trajectories.

\subsection{Commercial Tools and Academic Accessibility: A Systemic Crisis}
\label{subsec:commercial-academic-nexus}

Our findings reveal structural tethering of the research activity system to commercial tooling that systematically excludes BLV researchers. The convergence on general-purpose tools over specialized research software represents not preference but systemic exclusion from purpose-built infrastructure~\cite{aishwarya_performing_2022}, extending HCI critiques of academic dependence on proprietary technologies prioritizing profit over accessibility~\cite{Schonfeld_2019, deschamps_2023, Williams_2024}.

The "accessibility debt" we identified, the accumulated cognitive and emotional burden of anticipating tool failures, extends beyond immediate productivity impacts. P1's distrust reflects "learned technological helplessness"~\cite{harris2011towards}, where repeated failures condition researchers to anticipate inevitable exclusion. This burden compounds into internalized ableism~\cite{campbell_internalised_2009}, shaping decisions about methodologies, collaborations, and whether to remain in research. Most critically, research's visual-centric dissemination culture creates an insurmountable contradiction: BLV researchers must produce visual outputs they cannot independently evaluate. With less than 3\% of scholarly PDFs meeting basic accessibility standards~\cite{kumar_uncovering_2024, singh_chawla_accessibility_2024} and evaluation criteria privileging visual aesthetics over intellectual contribution, as P1's lost presentation award demonstrates, seemingly neutral academic practices encode ableist assumptions~\cite{brown_ableism_2018}.

The creative workarounds our participants developed, repurposing ChatGPT as a PDF reader, creating informal early-warning networks for accessibility changes, demonstrate the innovation under constraint frequently documented in HCI research~\cite{glazko_2023, adnin_2024}. However, we must resist the temptation to celebrate this resilience, as it risks normalizing the very exclusion that necessitated such workarounds, perpetuating epistemic violence against disabled researchers~\cite{ymous_i_2020}. As P3 noted, qualitative analysis tools should make text data analysis trivial; their inaccessibility represents egregious failure.

The parallel development of identical workarounds across institutions, multiple participants independently using Word for qualitative coding, exposes systematic market failure. Research tool developers, operating within capitalist frameworks prioritizing profitable features over accessibility~\cite{Schonfeld_2019, Williams_2024}, have effectively abandoned BLV researchers, forcing them into makeshift solutions that limit methodological options and research scope. Corporations' responses, NVivo's year-long non-response, perfunctory "we'll look into it" replies, reveal the complete absence of BLV researchers from their imagined user base, contradicting universal design and participatory design principles central to HCI~\cite{van_der_velden_participatory_2021}.

\subsubsection{The AI Paradox: Promise, Peril, and Verification Labor}

AI's emergence as an accessibility mediator offers both promise and peril. While providing unprecedented support for document summarization and visual description, it raises concerns about hallucination, data privacy, research integrity, and sustainability. Hallucination is especially problematic because BLV researchers cannot easily verify AI-generated visual outputs, creating what we term the \textit{verification paradox}: the accommodation enabling access simultaneously undermines verification ability.

Participants described labor-intensive verification strategies: cross-referencing AI outputs with multiple tools, seeking sighted colleague confirmation, and running queries through different systems. P2 acknowledged: "\textit{by AI hallucinating, [I] probably [will] miss things like some visual things}." P13 elaborated: "\textit{with AI [and] the problems with hallucination, it's almost going to take just as much research to go through to fact check everything}."

These tensions align with emerging HCI research. Glazko et al.~\cite{glazko_2023} found GenAI least useful when verification required the same accommodation it provided. Adnin and Das~\cite{adnin_2024} revealed that blind users develop sophisticated but labor-intensive verification strategies, often seeking sighted assistance or forgoing verification based on context, stakes, verifiability, and believability. For BLV researchers, this creates a troubling paradox: AI tools promise independence yet may require more human intervention to verify accuracy, transforming accessibility aids into potential vectors of misinformation where accuracy is paramount. This exemplifies the broader crisis at the intersection of commercial technology and academic accessibility. Industry pursues complex innovations while failing at basic accessibility, revealing how profit-driven priorities exclude disabled users~\cite{Williams_2024}. \textbf{The research community must move beyond celebrating individual workarounds to demand systemic accessibility from tool developers. Academic institutions must critically examine their structural dependence on commercial tooling and actively invest in accessible research infrastructure.}

\subsubsection{Re-Imagining Research Workflows: Adaptation vs. Transformation}

Our findings raise fundamental interaction design questions: should we retrofit existing tools for accessibility or design new tools aligned with BLV researchers' workflows? Participants' experiences suggest both approaches have limitations. As P4 noted, even "technically accessible" tools often require convoluted workflows, undermining efficiency. Yet purpose-built tools risk creating segregated workflows, isolating BLV researchers from collaborative environments.

We argue for a middle path: universal design principles~\cite{story_universal_design_1998, van_der_velden_participatory_2021} creating tools supporting multiple interaction modalities from inception, treating visual and non-visual interaction as equally first-class citizens, with participatory BLV researcher input throughout design. Such tools must integrate seamlessly with existing research ecosystems: standard file formats, cross-tool collaboration, and institutional infrastructure compatibility. \textbf{Tool developers should prioritize accessibility-first design, including BLV researchers from initial conception, ensuring diverse interaction modalities without forcing isolated workflows.}

\subsection{Ableist Assumptions and Epistemic Marginalization}
\label{subsec:ableism-and-epistemic-injustice}

Our findings reveal how ableist assumptions pervade collaborative dynamics and evaluation systems, creating compounding epistemic marginalization across multiple dimensions. From labor distribution to excellence recognition, these assumptions systematically disadvantage BLV researchers through both explicit exclusion and implicit bias.

\subsubsection{Interdependence, Autonomy, and Collaborative Constraints}

Collaborative configurations BLV researchers navigate complicate narratives around teamwork, revealing how accessibility needs reshape team dynamics in ways that both enable and constrain professional development. P11's task redistribution, conducting interviews while sighted colleagues create visualizations, maintains productivity but may create skill silos, limiting advancement and constraining what counts as "appropriate" contributions. This reflects power asymmetries in collaborative work documented throughout HCI literature~\cite{das_2019, akter_2025}.

This pattern exemplifies interdependence~\cite{bennett_interdependence_2018}: collaborative access strategies overcome immediate barriers while potentially creating new dependencies. While offering clear advantages, this approach risks confining BLV researchers to predetermined "accessible" tasks rather than supporting autonomous contribution choices. This extends beyond teams to institutional structures. When one partner in an academic couple is BLV, decisions about dual-career hiring, laboratory space, and institutional support must account for access needs affecting both careers, yet institutions rarely have frameworks addressing such interdependent requirements. These gaps disadvantage BLV researchers in hiring negotiations and limit career-crucial geographic mobility.

Credibility challenges suggest collaboration occurs against a backdrop of ableist assumptions. P9's "excellence imperative," working "twice or thrice as hard" to prove competence, transforms collaboration into a proving ground. This parallels experiences documented in workplace accessibility research~\cite{marathe_2025} and extends epistemic injustice research~\cite{scully_she_2018}: BLV researchers must constantly demonstrate competence while managing accessibility needs, experiencing what Dotson terms "testimonial smothering"~\cite{dotson_tracking_2011}, the silencing of marginalized voices through systematic credibility deficits leading to self-censorship.

\subsubsection{Visual Privilege and Epistemic Exclusion in Evaluation}

Epistemic marginalization extends into formal evaluation systems, privileging visual presentation over intellectual contribution. P1's lost presentation award, despite strong content, may demonstrate how evaluation criteria systematically disadvantage BLV researchers. This reflects not individual bias but structural discrimination embedded in promotion, grants, and publication decisions~\cite{brown_ableism_2018}.

As P1 noted, "\textit{visual elements dominate the slides and the slides make a very important part of a presentation}." When evaluation systems conflate medium with quality, they create what Scully describes as \textit{hermeneutical injustice}~\cite{scully_she_2018}: BLV researchers' contributions are undervalued because evaluative frameworks embody ableist assumptions about what recognizing excellence looks like. This bias becomes especially insidious given pervasive accessibility failures in publishing~\cite{wang_improving_2021, bigham_uninteresting_2016}, creating a vicious cycle where inaccessible dissemination reinforces visual privilege in evaluation.

\subsubsection{Toward Epistemic Justice: Dual Imperatives}

These compounding marginalizations demand fundamental reconceptualization of research participation and excellence. Research communities must recognize that BLV researchers require \textit{both} autonomy in choosing contributions \textit{and} evaluation systems decoupling intellectual contribution from visual aesthetics. This dual imperative, autonomy in participation and equity in evaluation, is essential for epistemic justice.

Alternative frameworks might emphasize contribution, rigor, and impact while de-emphasizing aesthetics: review processes evaluating content separately from presentation, or alternative formats such as rich audio presentations, interactive tactile diagrams, and narrative demonstrations that play to diverse strengths. Conference organizers might offer accessible slide design support or develop rubrics explicitly valuing content over aesthetics.

More fundamentally, evaluation must shift from treating accessibility as individual accommodation to recognizing it as a dimension of research quality itself. Just as methodological rigor and ethical practice are core evaluation criteria, output accessibility should be integral to scholarly communication. Papers with inaccessible figures, presentations with inadequate descriptions, and datasets with undocumented formats should be considered incomplete work, regardless of visual appeal.

Without both dimensions, efforts risk reproducing epistemic marginalization in new forms: either constraining BLV researchers to limited research roles or undervaluing their contributions through biased evaluation criteria. \textbf{Research communities must enable BLV researchers' autonomous participation across all activities while transforming evaluation systems to assess intellectual contribution independent of visual aesthetics.}

\subsection{Actionable Recommendations for Research Communities}
\label{subsec:concrete-actions}

Drawing on our findings and participants' explicit requests, we offer concrete actions for the HCI and broader research community, emerging from both direct advocacy and our systemic contradiction analysis.

\subsubsection{For Tool Developers and Technology Companies}

\textbf{Accessibility-first design must become the default, not an add-on.} This requires including BLV researchers in initial design phases, not just testing. As P4 noted, companies test "\textit{for the outward-facing part of their platform}" but ignore the "programmer side," making assumptions about who creates versus consumes research. Participatory design~\cite{van_der_velden_participatory_2021} offers established frameworks for centering disabled users' experiences from inception.

Equally critical is tooling \textit{stability}: ensure accessibility remains consistent across versions through stability options, backwards compatibility, and mandatory accessibility testing before updates. The "update anxiety" participants described, where each update threatens established workflows, stems from treating accessibility as an afterthought rather than core functionality. Tool developers must establish transparent communication channels for accessibility issues with committed timelines and public roadmaps. Year-long silences and vague promises erode trust and prevent planning.

\subsubsection{For Research Institutions and Funding Agencies}

\textbf{Institutions must recognize and support ability-informed research pathways.} We must acknowledge the hidden labor BLV researchers expend to maintain accessible workflows and provide dedicated accessibility support as standard infrastructure: funding for technology, technical support, and explicit recognition of accessibility labor in promotion decisions.

Institutions should develop frameworks for interdependent access in dual-career hiring and collaborations. When one partner is BLV, both careers are affected by accessibility considerations, yet institutions rarely account for this in negotiations, space allocation, or support structures. Funding agencies should require accessibility plans in grant applications, not just for recruitment but for tools and dissemination, and provide supplemental funding for accessibility needs.

\subsubsection{For Conference Organizers and Publishers}

\textbf{We push for fundamental re-evaluation of evaluation criteria.} As P12 asked, "how do I know if it looks good visually?" This question points toward developing tools that convey visual effectiveness through non-visual channels: AI-assisted evaluation, standardized templates, alternative formats de-emphasizing visual design while maintaining intellectual standards. Conference organizers should offer accessible presentation support, including slide review, technical assistance, and templates, and develop rubrics explicitly prioritizing content over visual polish. Review processes should evaluate content separately from presentation, recognizing that compelling research communicates through diverse modalities.

Publishers must enforce accessibility standards, treating inaccessible figures and tables as incomplete submissions requiring revision, just as they would for methodological flaws or ethical concerns. This represents a fundamental shift: from accessibility as authors' sole responsibility to accessibility as a community-wide commitment and an HCI design challenge requiring collective attention.

\subsubsection{For the Broader Research Community}

\textbf{The research community must actively dismantle ableist assumptions pervading research culture.} This requires: recognizing "accessibility debt" as invisible labor deserving institutional support; creating space for BLV researchers' autonomous contributions across all research activities; advocating for structural infrastructure changes rather than expecting continuous individual workarounds; and recognizing that AI progress does not excuse persistent basic accessibility failures.

Most critically, the community must \textit{listen to and center BLV researchers' expertise} about their needs. BLV researchers are not problems to solve but knowledgeable collaborators in re-imagining accessible, equitable research systems. Their experiences reveal fundamental flaws in how research infrastructure operates, offering invaluable insights for transformation.

\subsection{Limitations and Future Directions}
\label{subsec:limitations_and_future_directions}

Our participant pool represents a targeted sample rather than a comprehensive cross-section, reflecting broader challenges in disabled population research where shared researcher-participant identity often enhances recruitment and rapport~\cite{emara_talking_2025}. Our sampling strategy may have introduced selection bias~\cite{winship_models_1992}: recruitment through advocacy organizations may capture more resource-aware researchers than those struggling in isolation. Our cross-sectional design captures a snapshot that may evolve as technologies develop. Longitudinal studies could reveal how BLV researchers' preferences and challenges shift over time~\cite{fichten_higher_2020}, particularly as research increasingly emphasizes data-driven methodologies and new visualization paradigms.

We focused primarily on \textit{digital} tool accessibility in desk-based activities. Research ecology extends far beyond this to laboratory equipment, field sites, conference venues, and physical infrastructure. These material and spatial aspects warrant separate investigation, as do discipline-specific challenges in experimental sciences, field work, or clinical research. Future work should examine how accessibility challenges vary across disciplines, career stages, and contexts, while investigating the effectiveness of different intervention strategies. We acknowledge limitations of grouping research tools into specific categories~\cite{creswell_research_2017}; recent work has explored variability in usability and accessibility of digital collaborative tools among BLV workers~\cite{akter_2025}. Future work may consider alternate classifications capturing such nuance.

While surfacing BLV researchers' experiences, we recognize that other disabled researchers face intersecting and distinct barriers warranting dedicated investigation. Building on living disability theory frameworks~\cite{hofmann_living_2020}, future research should explore how different disabilities shape research participation and how systemic solutions might address multiple forms of exclusion simultaneously. These limitations highlight this area's nascent state and underscore the urgent need for sustained HCI investigation. The systemic contradictions we identified are not peripheral concerns but fundamental challenges to creating equitable research systems that benefit from the full diversity of human experience and capability.
\section{Conclusion}
\label{sec:conclusion}

We designed an explanatory sequential mixed-methods study to explore BLV researchers’ workflows and processes. Our findings reveal that BLV researchers face pervasive accessibility barriers across the research lifecycle, from literature review platforms to specialized analysis software to journal submission systems, necessitating extensive workarounds, tool substitutions, and reliance on sighted collaborators that fundamentally reshape how they conduct research. Despite these systemic challenges, participants developed alternative workflows, leveraging emerging technologies like AI for transcription and image description, and strategically redistributing collaborative tasks based on accessibility constraints. However, these adaptations often come at significant costs, requiring additional time, limiting tool choices, affecting networking opportunities, and sometimes forcing researchers to abandon certain ideas entirely. Our findings highlight stark variations in institutional support, with some researchers benefiting from comprehensive disability services while others navigate accessibility challenges largely independently. Participants emphasized that true accessibility requires moving beyond retrofitted solutions toward universal design principles embedded from the ground up across the entire research activity system. We advocate for designing and developing evidence-based research tools in partnership with BLV researchers, particularly data analysis and visualization tools, and creating collaborative frameworks that leverage the innovative workarounds and strategies documented in this study to benefit the broader research community. Additionally, as AI tools become more prevalent, we must remain informed on how these technologies can be thoughtfully integrated to enhance accessibility without replacing the critical analytical and creative work that defines quality research.


  \begin{acks}

We thank all of our participants for their time in completing this study, as we look towards making research an accessible community of practice for all. This study was funded by the last author's faculty startup fund. 

\end{acks}

\bibliographystyle{ACM-Reference-Format}
\bibliography{a11y_framework,references}

@inproceedings{mankoffDisabilityStudiesSource2010,
  title = {Disability Studies as a Source of Critical Inquiry for the Field of Assistive Technology},
  booktitle = {Proceedings of the 12th International {{ACM SIGACCESS}} Conference on {{Computers}} and Accessibility},
  author = {Mankoff, Jennifer and Hayes, Gillian R. and Kasnitz, Devva},
  year = {2010},
  month = oct,
  series = {{{ASSETS}} '10},
  pages = {3--10},
  publisher = {Association for Computing Machinery},
  address = {New York, NY, USA},
  doi = {10.1145/1878803.1878807},
  urldate = {2022-11-06},
  isbn = {978-1-60558-881-0}
}

@inproceedings{sharifShouldSayDisabled2022,
  title = {Should {{I Say}} ``{{Disabled People}}'' or ``{{People}} with {{Disabilities}}''? {{Language Preferences}} of {{Disabled People Between Identity-}} and {{Person-First Language}}},
  shorttitle = {Should {{I Say}} ``{{Disabled People}}'' or ``{{People}} with {{Disabilities}}''?},
  booktitle = {Proceedings of the 24th {{International ACM SIGACCESS Conference}} on {{Computers}} and {{Accessibility}}},
  author = {Sharif, Ather and McCall, Aedan Liam and Bolante, Kianna Roces},
  year = {2022},
  month = oct,
  series = {{{ASSETS}} '22},
  pages = {1--18},
  publisher = {Association for Computing Machinery},
  address = {New York, NY, USA},
  doi = {10.1145/3517428.3544813},
  urldate = {2024-06-23},
  isbn = {978-1-4503-9258-7}
}

@inproceedings{aishwarya_performing_2022,
  title = {Performing {{Qualitative Data Analysis}} as a {{Blind Researcher}}: {{Challenges}}, {{Workarounds}} and {{Design Recommendations}}},
  shorttitle = {Performing {{Qualitative Data Analysis}} as a {{Blind Researcher}}},
  booktitle = {Proceedings of the 24th {{International ACM SIGACCESS Conference}} on {{Computers}} and {{Accessibility}}},
  author = {Aishwarya, O.},
  year = {2022},
  month = oct,
  series = {{{ASSETS}} '22},
  pages = {1--4},
  publisher = {Association for Computing Machinery},
  address = {New York, NY, USA},
  doi = {10.1145/3517428.3551356},
  urldate = {2025-04-18},
  isbn = {978-1-4503-9258-7},
  
}

@misc{al-jadir10DisabledScientists2024,
  title = {10 {{Disabled Scientists To Discover}}},
  author = {{Al-Jadir}, Raya},
  year = {2024},
  month = may,
  journal = {Disability Horizons},
  urldate = {2025-09-06},
  langid = {british},
  
}

@inproceedings{bennett_interdependence_2018,
  title = {Interdependence as a {{Frame}} for {{Assistive Technology Research}} and {{Design}}},
  booktitle = {Proceedings of the 20th {{International ACM SIGACCESS Conference}} on {{Computers}} and {{Accessibility}}},
  author = {Bennett, Cynthia L. and Brady, Erin and Branham, Stacy M.},
  year = {2018},
  month = oct,
  series = {{{ASSETS}} '18},
  pages = {161--173},
  publisher = {Association for Computing Machinery},
  address = {New York, NY, USA},
  doi = {10.1145/3234695.3236348},
  urldate = {2025-04-18},
  isbn = {978-1-4503-5650-3},
  
}

@inproceedings{bigham_uninteresting_2016,
  title = {An {{Uninteresting Tour Through Why Our Research Papers Aren}}'t {{Accessible}}},
  booktitle = {Proceedings of the 2016 {{CHI Conference Extended Abstracts}} on {{Human Factors}} in {{Computing Systems}}},
  author = {Bigham, Jeffrey P. and Brady, Erin L. and Gleason, Cole and Guo, Anhong and Shamma, David A.},
  year = {2016},
  month = may,
  series = {{{CHI EA}} '16},
  pages = {621--631},
  publisher = {Association for Computing Machinery},
  address = {New York, NY, USA},
  doi = {10.1145/2851581.2892588},
  urldate = {2025-05-14},
  isbn = {978-1-4503-4082-3},

}

@article{brown_ableism_2018,
  title = {Ableism in Academia: Where Are the Disabled and Ill Academics?},
  shorttitle = {Ableism in Academia},
  author = {Brown, Nicole and {and Leigh}, Jennifer},
  year = {2018},
  month = jul,
  journal = {Disability \& Society},
  volume = {33},
  number = {6},
  pages = {985--989},
  publisher = {Routledge},
  issn = {0968-7599},
  doi = {10.1080/09687599.2018.1455627},
  urldate = {2025-04-18},

}

@incollection{campbell_internalised_2009,
  title = {Internalised {{Ableism}}: {{The Tyranny Within}}},
  shorttitle = {Internalised {{Ableism}}},
  booktitle = {Contours of {{Ableism}}: {{The Production}} of {{Disability}} and {{Abledness}}},
  author = {Campbell, Fiona Kumari},
  editor = {Campbell, Fiona Kumari},
  year = {2009},
  pages = {16--29},
  publisher = {Palgrave Macmillan UK},
  address = {London},
  doi = {10.1057/9780230245181_2},
  urldate = {2025-04-18},
  isbn = {978-0-230-24518-1},
  langid = {english}
}

@article{castrodale_youre_2015,
  title = {``{{You}}'re Such a Good Friend'': {{A}} Woven Autoethnographic Narrative Discussion of Disability and Friendship in {{Higher Education}}},
  shorttitle = {``{{You}}'re Such a Good Friend''},
  author = {Castrodale, Mark Anthony and Zingaro, Daniel},
  year = {2015},
  month = feb,
  journal = {Disability Studies Quarterly},
  volume = {35},
  number = {1},
  issn = {2159-8371},
  doi = {10.18061/dsq.v35i1.3762},
  urldate = {2025-04-22},
  copyright = {Copyright (c) 2015 Mark Anthony Castrodale, Daniel Zingaro},
  langid = {english}
}

@misc{crd_blv_report_2025,
  title = {Compilation and {{Expansion A}}: {{Statistics}} on the {{Blind}} and {{Low Vision Population}} - {{Compendium}} (2025)},
  shorttitle = {Compilation and {{Expansion A}}},
  author = {{Center for Research on Disability}},
  year = {2025},
  month = mar,
  journal = {Center for Research on Disability},
  urldate = {2025-06-27},
  howpublished = {https://www.researchondisability.org/annual-disability-statistics-collection/2025-compendium-table-contents/compilation-expansion-statistics-blind-low-vision-population-compendium-2025},
  langid = {english},
  
}

@book{creswell_research_2017,
  title = {Research {{Design}}: {{Qualitative}}, {{Quantitative}}, and {{Mixed Methods Approaches}}},
  shorttitle = {Research {{Design}}},
  author = {Creswell, John W. and Creswell, J. David},
  year = {2017},
  month = dec,
  publisher = {SAGE Publications},
  googlebooks = {335ZDwAAQBAJ},
  isbn = {978-1-5063-8669-0},
  langid = {english}
}

@article{dotson_tracking_2011,
  title = {Tracking {{Epistemic Violence}}, {{Tracking Practices}} of {{Silencing}}},
  author = {Dotson, Kristie},
  year = {2011},
  journal = {Hypatia},
  volume = {26},
  number = {2},
  pages = {236--257},
  issn = {1527-2001},
  doi = {10.1111/j.1527-2001.2011.01177.x},
  urldate = {2025-04-22},
  copyright = {{\copyright} by Hypatia, Inc.},
  langid = {english}
}

@article{emara_talking_2025,
  title = {``{{Talking}} the {{Same Language}}'': {{The Influence}} of {{Sharing}} a {{Visual Impairment Identity Between Researchers}} and {{Participants}} on {{Enhancing Participant Recruitment}} and {{Fostering Rapport During Interviews With Blind Individuals}}},
  shorttitle = {``{{Talking}} the {{Same Language}}''},
  author = {Emara, Ibrahim},
  year = {2025},
  month = apr,
  journal = {International Journal of Qualitative Methods},
  volume = {24},
  pages = {16094069251320858},
  publisher = {SAGE Publications Inc},
  issn = {1609-4069},
  doi = {10.1177/16094069251320858},
  urldate = {2025-05-15},
  langid = {english},
  
}

@article{greenvall_influence_2021,
  title = {The {{Influence}} of a {{Blind Professor}} in a {{Bioengineering Course}}},
  author = {Greenvall, Benjamin R. and Tiano, Amanda L. and Chandani, Anjali and Minkara, Mona S.},
  year = {2021},
  month = jul,
  journal = {Biomedical Engineering Education},
  volume = {1},
  number = {2},
  pages = {245--258},
  issn = {2730-5945},
  doi = {10.1007/s43683-021-00052-1},
  urldate = {2025-05-14},
  langid = {english},
  
}

@article{hartmann_disability_2019,
  title = {Disability Inclusion Enhances Science},
  author = {Hartmann, Aaron C.},
  year = {2019},
  month = nov,
  journal = {Science},
  volume = {366},
  number = {6466},
  pages = {698--698},
  publisher = {American Association for the Advancement of Science},
  doi = {10.1126/science.aaz0271},
  urldate = {2025-05-14},
  
}

@inproceedings{hofmann_living_2020,
  title = {Living {{Disability Theory}}: {{Reflections}} on {{Access}}, {{Research}}, and {{Design}}},
  shorttitle = {Living {{Disability Theory}}},
  booktitle = {Proceedings of the 22nd {{International ACM SIGACCESS Conference}} on {{Computers}} and {{Accessibility}}},
  author = {Hofmann, Megan and Kasnitz, Devva and Mankoff, Jennifer and Bennett, Cynthia L},
  year = {2020},
  month = oct,
  series = {{{ASSETS}} '20},
  pages = {1--13},
  publisher = {Association for Computing Machinery},
  address = {New York, NY, USA},
  doi = {10.1145/3373625.3416996},
  urldate = {2025-04-22},
  isbn = {978-1-4503-7103-2},
  
}

@inproceedings{jain_autoethnography_2019,
  title = {Autoethnography of a {{Hard}} of {{Hearing Traveler}}},
  booktitle = {Proceedings of the 21st {{International ACM SIGACCESS Conference}} on {{Computers}} and {{Accessibility}}},
  author = {Jain, Dhruv and Desjardins, Audrey and Findlater, Leah and Froehlich, Jon E.},
  year = {2019},
  month = oct,
  series = {{{ASSETS}} '19},
  pages = {236--248},
  publisher = {Association for Computing Machinery},
  address = {New York, NY, USA},
  doi = {10.1145/3308561.3353800},
  urldate = {2025-04-22},
  isbn = {978-1-4503-6676-2}
}

@inproceedings{jain_navigating_2020,
  title = {Navigating {{Graduate School}} with a {{Disability}}},
  booktitle = {Proceedings of the 22nd {{International ACM SIGACCESS Conference}} on {{Computers}} and {{Accessibility}}},
  author = {Jain, Dhruv and Potluri, Venkatesh and Sharif, Ather},
  year = {2020},
  month = oct,
  series = {{{ASSETS}} '20},
  pages = {1--11},
  publisher = {Association for Computing Machinery},
  address = {New York, NY, USA},
  doi = {10.1145/3373625.3416986},
  urldate = {2025-04-22},
  isbn = {978-1-4503-7103-2},
  
}

@article{katsnelson_these_2023,
  title = {These Tools Help Visually Impaired Scientists Read Data and Journals},
  author = {Katsnelson, Alla},
  year = {2023},
  month = mar,
  journal = {Nature},
  volume = {615},
  number = {7951},
  pages = {362--363},
  publisher = {Nature Publishing Group},
  doi = {10.1038/d41586-023-00645-6},
  urldate = {2025-05-14},
  copyright = {2023 Springer Nature Limited},
  langid = {english}
}

@inproceedings{kumar_uncovering_2024,
  title = {Uncovering the {{New Accessibility Crisis}} in {{Scholarly PDFs}}: {{Publishing Model}} and {{Platform Changes Contribute}} to {{Declining Scholarly Document Accessibility}} in the {{Last Decade}}},
  shorttitle = {Uncovering the {{New Accessibility Crisis}} in {{Scholarly PDFs}}},
  booktitle = {Proceedings of the 26th {{International ACM SIGACCESS Conference}} on {{Computers}} and {{Accessibility}}},
  author = {Kumar, Anukriti and Wang, Lucy Lu},
  year = {2024},
  month = oct,
  series = {{{ASSETS}} '24},
  pages = {1--16},
  publisher = {Association for Computing Machinery},
  address = {New York, NY, USA},
  doi = {10.1145/3663548.3675634},
  urldate = {2025-05-14},
  isbn = {979-8-4007-0677-6},
  
}

@inproceedings{manzoor_assistive_2018,
  title = {Assistive {{Debugging}} to {{Support Accessible Latex Based Document Authoring}}},
  booktitle = {Proceedings of the 20th {{International ACM SIGACCESS Conference}} on {{Computers}} and {{Accessibility}}},
  author = {Manzoor, Ahtsham and Parvez, Murayyiam and Shahid, Suleman and Karim, Asim},
  year = {2018},
  month = oct,
  series = {{{ASSETS}} '18},
  pages = {432--434},
  publisher = {Association for Computing Machinery},
  address = {New York, NY, USA},
  doi = {10.1145/3234695.3241013},
  urldate = {2025-04-18},
  isbn = {978-1-4503-5650-3},
  
}

@article{mcruer_proliferating_2014,
  title = {Proliferating {{Cripistemologies}}: {{A Virtual Roundtable}}},
  shorttitle = {Proliferating {{Cripistemologies}}},
  author = {McRuer, Robert and Johnson, Merri Lisa},
  year = {2014},
  journal = {Journal of Literary \& Cultural Disability Studies},
  volume = {8},
  number = {2},
  pages = {149--169},
  publisher = {Liverpool University Press},
  issn = {1757-6466},
  urldate = {2025-04-22}
}

@article{minkara_implementation_2015,
  title = {Implementation of {{Protocols To Enable Doctoral Training}} in {{Physical}} and {{Computational Chemistry}} of a {{Blind Graduate Student}}},
  author = {Minkara, Mona S. and Weaver, Michael N. and Gorske, Jim and Bowers, Clifford R. and Merz, Kenneth M. Jr.},
  year = {2015},
  month = aug,
  journal = {Journal of Chemical Education},
  volume = {92},
  number = {8},
  pages = {1280--1283},
  publisher = {American Chemical Society},
  issn = {0021-9584},
  doi = {10.1021/ed5009552},
  urldate = {2025-05-14},
  
}

@misc{MonaMinkaraBlind,
  title = {Mona {{Minkara}} {\textbar} {{Blind Scientist Philosophy}}},
  author = {Minkara, Mona S.},
  year = {2024},
  urldate = {2025-09-06},
  howpublished = {https://monaminkara.com/unseen-advantage\#gsc.tab=0},
  
}

@article{rizzo_accessible_2024,
  title = {Accessible Scientific Conferences for Blind and Low Vision Professionals and Researchers: A Necessary Step for Achieving {{STEMM}} Equity},
  shorttitle = {Accessible Scientific Conferences for Blind and Low Vision Professionals and Researchers},
  year = {2024},
  author = {Rizzo, John-Ross and , Penny, Rosenblum and , Christina, Samuel and , Walter, Wittich and , Natalina, Martiniello and , Mahya, Beheshti and , Aaron, Johnson and , Yueh-Hsub, Wu and , Mahadeo, Sukhai and {and Bonnielin}, Swenor},
  journal = {Disability \& Society},
  volume = {0},
  number = {0},
  pages = {1--7},
  publisher = {Routledge},
  issn = {0968-7599},
  doi = {10.1080/09687599.2024.2412269},
  urldate = {2025-05-14},
 
}

@article{sahtout_how_2020,
  title = {How Science Should Support Researchers with Visual Impairments},
  author = {Sahtout, Naheda},
  year = {2020},
  month = sep,
  journal = {Nature},
  publisher = {Nature Publishing Group},
  doi = {10.1038/d41586-020-02627-4},
  urldate = {2025-05-14},
  copyright = {2020 Springer Nature Limited},
  langid = {english},
}

@article{winship_models_1992,
  title={Models for sample selection bias},
  author={Winship, Christopher and Mare, Robert D},
  journal={Annual review of sociology},
  volume={18},
  number={1},
  pages={327--350},
  year={1992},
  publisher={Annual Reviews 4139 El Camino Way, PO Box 10139, Palo Alto, CA 94303-0139, USA}
}

@inproceedings{schmitt-koopmann_more_2025,
  title = {Towards {{More Accessible Scientific PDFs}} for {{People}} with {{Visual Impairments}}: {{Step-by-Step PDF Remediation}} to {{Improve Tag Accuracy}}},
  shorttitle = {Towards {{More Accessible Scientific PDFs}} for {{People}} with {{Visual Impairments}}},
  booktitle = {Proceedings of the 2025 {{CHI Conference}} on {{Human Factors}} in {{Computing Systems}}},
  author = {{Schmitt-Koopmann}, Felix Maximilian and Huang, Elaine May and Hutter, Hans-Peter and Darvishy, Alireza},
  year = {2025},
  month = apr,
  series = {{{CHI}} '25},
  pages = {1--16},
  publisher = {Association for Computing Machinery},
  address = {New York, NY, USA},
  doi = {10.1145/3706598.3713084},
  urldate = {2025-05-14},
  isbn = {979-8-4007-1394-1},
  
}

@misc{ScienceGoldenInterviews,
  title = {Science {{Is Golden}}: {{Interviews}} with {{Four Scientists Who Are Visually Impaired}}},
  shorttitle = {Science {{Is Golden}}},
  author = {{American Foundation for the Blind}},
  year = {2005},
  journal = {The American Foundation for the Blind},
  urldate = {2025-09-06},
  howpublished = {https://www.afb.org/aw/6/1/14656},
  langid = {english},
  
}

@article{scully_she_2018,
  title = {From ``{{She Would Say That}}, {{Wouldn}}'t {{She}}?'' To ``{{Does She Take Sugar}}?'' {{Epistemic Injustice}} and {{Disability}}},
  shorttitle = {From ``{{She Would Say That}}, {{Wouldn}}'t {{She}}?},
  author = {Scully, Jackie Leach},
  year = {2018},
  journal = {IJFAB: International Journal of Feminist Approaches to Bioethics},
  volume = {11},
  number = {1},
  pages = {106--124},
  publisher = {University of Toronto Press},
  issn = {1937-4577},
  urldate = {2025-04-22}
}

@article{singh_chawla_accessibility_2024,
  title = {Accessibility Worsens for Blind and Low-Vision Readers of Academic {{PDFs}}},
  author = {Singh Chawla, Dalmeet},
  year = {2024},
  month = dec,
  journal = {Nature},
  publisher = {Nature Publishing Group},
  issn = {1476-4687},
  doi = {10.1038/d41586-024-03953-7},
  urldate = {2025-05-14},
  copyright = {2024 Springer Nature Limited},
  langid = {english},
  
}

@misc{unesco_2021,
  title = {Statistics and Resources {\textbar} 2021 {{Science Report}}},
  author = {{UNESCO}},
  year = {2021},
  month = may,
  urldate = {2025-06-27},
  howpublished = {https://www.unesco.org/reports/science/2021/en/statistics},
  langid = {english},
  
}

@article{swartzScienceValueDiversity2019,
  title = {The {{Science}} and {{Value}} of {{Diversity}}: {{Closing}} the {{Gaps}} in {{Our Understanding}} of {{Inclusion}} and {{Diversity}}},
  shorttitle = {The {{Science}} and {{Value}} of {{Diversity}}},
  author = {Swartz, Talia H and Palermo, Ann-Gel S and Masur, Sandra K and Aberg, Judith A},
  year = {2019},
  month = sep,
  journal = {The Journal of Infectious Diseases},
  volume = {220},
  number = {Suppl 2},
  pages = {S33-S41},
  issn = {0022-1899},
  doi = {10.1093/infdis/jiz174},
  urldate = {2025-09-06},
  pmcid = {PMC6701939},
  pmid = {31430380},
  
}

@misc{trager_visionary_2024,
  title = {Visionary Chemistry Is Making Labs Accessible to Blind Students and Researchers},
  year = {2024},
  author = {Trager, Rebecca},
  journal = {Chemistry World},
  urldate = {2025-05-14},
  howpublished = {https://www.chemistryworld.com/careers/visionary-chemistry-is-making-labs-accessible-to-blind-students-and-researchers/4018758.article},
  langid = {english}
}

@incollection{van_der_velden_participatory_2021,
  title = {Participatory {{Design}} and {{Design}} for {{Values}}},
  booktitle = {Handbook of {{Ethics}}, {{Values}}, and {{Technological Design}}: {{Sources}}, {{Theory}}, {{Values}} and {{Application Domains}}},
  author = {{van der Velden}, Maja and M{\"o}rtberg, Christina},
  editor = {{van den Hoven}, Jeroen and Vermaas, Pieter E. and {van de Poel}, Ibo},
  year = {2021},
  pages = {1--22},
  publisher = {Springer Netherlands},
  address = {Dordrecht},
  doi = {10.1007/978-94-007-6994-6_33-1},
  urldate = {2025-04-22},
  isbn = {978-94-007-6994-6},
  langid = {english}
}

@misc{wang_improving_2021,
  title = {Improving the {{Accessibility}} of {{Scientific Documents}}: {{Current State}}, {{User Needs}}, and a {{System Solution}} to {{Enhance Scientific PDF Accessibility}} for {{Blind}} and {{Low Vision Users}}},
  shorttitle = {Improving the {{Accessibility}} of {{Scientific Documents}}},
  author = {Wang, Lucy Lu and Cachola, Isabel and Bragg, Jonathan and Cheng, Evie Yu-Yen and Haupt, Chelsea and Latzke, Matt and Kuehl, Bailey and van Zuylen, Madeleine and Wagner, Linda and Weld, Daniel S.},
  year = {2021},
  month = apr,
  number = {arXiv:2105.00076},
  eprint = {2105.00076},
  primaryclass = {cs},
  publisher = {arXiv},
  doi = {10.48550/arXiv.2105.00076},
  urldate = {2025-04-18},
  archiveprefix = {arXiv},
  
}

@misc{who_2023_disability_report,
  title = {World {{Report}} on {{Disability}}},
  author = {World Health Organization},
  year = {2023},
  month = feb,
  urldate = {2025-04-22},
  howpublished = {https://www.who.int/teams/noncommunicable-diseases/sensory-functions-disability-and-rehabilitation/world-report-on-disability},
  langid = {english}
}

@inproceedings{ymous_i_2020,
  title = {"{{I}} Am Just Terrified of My Future" --- {{Epistemic Violence}} in {{Disability Related Technology Research}}},
  booktitle = {Extended {{Abstracts}} of the 2020 {{CHI Conference}} on {{Human Factors}} in {{Computing Systems}}},
  author = {Ymous, Anon and Spiel, Katta and Keyes, Os and Williams, Rua M. and Good, Judith and Hornecker, Eva and Bennett, Cynthia L.},
  year = {2020},
  month = apr,
  series = {{{CHI EA}} '20},
  pages = {1--16},
  publisher = {Association for Computing Machinery},
  address = {New York, NY, USA},
  doi = {10.1145/3334480.3381828},
  urldate = {2025-04-22},
  isbn = {978-1-4503-6819-3}
}

@incollection{engestrom_2014_activity,
  title={Activity theory and learning at work},
  author={Engestr{\"o}m, Yrj{\"o}},
  booktitle={T{\"a}tigkeit-Aneignung-Bildung: Positionierungen zwischen Virtualit{\"a}t und Gegenst{\"a}ndlichkeit},
  pages={67--96},
  year={2014},
  publisher={Springer}
}

@book{kaptelinin_2012_activity,
  title={Activity theory in HCI: Fundamentals and reflections},
  author={Kaptelinin, Victor and Nardi, Bonnie A},
  volume={13},
  year={2012},
  publisher={Morgan \& Claypool Publishers}
}

@book{vygotsky_1934_thought,
  author    = {Vygotsky, Lev Semenovich},
  title     = {Thought and Language},
  year      = {1934},
  publisher = {MIT Press},
  address   = {Cambridge, MA},
  note      = {Originally published in Russian as "Myshleniye i rech'"}
}

@book{leontiev_1973_problems,
  author    = {Leontiev, Alexei Nikolaevich},
  title     = {Problems of the Development of the Mind},
  year      = {1973},
  publisher = {Progress Publishers},
  address   = {Moscow},
  note      = {Originally published in Russian 1959}
}

@book{activity_theory_2012,
  title = {Activity {{Theory}}},
  author = {Kaptelinin, Victor and Nardi, Bonnie A},
  volume = {13},
  year = {2012},
  publisher = {Morgan \& Claypool Publishers},
  langid = {english},
}

@misc{activity_theory_overview,
  title = {Activity {{Theory}} - an Overview {\textbar} {{ScienceDirect Topics}}},
  author = {{ScienceDirect}},
  year = {2024},
  urldate = {2025-09-06},
  howpublished = {https://www.sciencedirect.com/topics/psychology/activity-theory}
}

@misc{concept_mediation_activity_2021,
  title = {The {{Concept}} of {{Mediation}} [{{Activity Theory}}]},
  author = {Engestr{\"o}m, Yrj{\"o}},
  year = {2021},
  month = oct,
  journal = {Activity Analysis Center},
  urldate = {2025-09-06},
  howpublished = {https://www.activityanalysis.net/the-concept-of-mediation/},
  langid = {english},
  
}

@misc{hierarchy_human_activity_2021,
  title = {The {{Hierarchy}} of {{Human Activity}} [{{Activity Theory}}]},
  author = {Engestr{\"o}m, Yrj{\"o}},
  year = {2021},
  month = oct,
  journal = {Activity Analysis Center},
  urldate = {2025-09-06},
  howpublished = {https://www.activityanalysis.net/the-hierarchy-of-human-activity/},
  langid = {english},
}

@misc{learning_by_expanding,
  title = {{{LEARNING BY EXPANDING}}},
  author = {Engestr{\"o}m, Yrj{\"o}},
  year = {1973},
  urldate = {2025-09-06},
  howpublished = {https://lchc.ucsd.edu/mca/Paper/Engestrom/expanding/intro.htm},
  langid = {english},
}

@misc{soviet_psychology_cw,
  title = {Soviet {{Psychology}}: {{C}}.{{W}}. {{Tolman}} on {{Leontiev}}},
  author = {Tolman, C. W.},
  year = {1973},
  urldate = {2025-09-06},
  howpublished = {https://www.marxists.org/archive/leontev/comment/tolman.htm},
}

@article{kaptelinin_activity_theory_framework_2018,
  title = {Activity {{Theory}} as a {{Framework}} for {{Human-Technology Interaction Research}}},
  author = {Kaptelinin, Victor and Nardi, Bonnie},
  year = {2018},
  month = jan,
  journal = {Mind, Culture, and Activity},
  volume = {25},
  number = {1},
  pages = {3--5},
  publisher = {Routledge},
  issn = {1074-9039},
  doi = {10.1080/10749039.2017.1393089},
  urldate = {2025-09-06},
  
}

@book{olson_ways_of_knowing_in_hci_2014,
  title={Ways of Knowing in HCI},
  author={Olson, Judith S and Kellogg, Wendy A},
  volume={2},
  year={2014},
  langid={english},
  publisher={Springer}
}

@misc{qualtrics,
    author = {Qualtrics},
    url = {https://www.qualtrics.com},
    year = {2025}
}

@article{goodman_snowball_1961,
  title={Snowball sampling},
  author={Goodman, Leo A},
  journal={The annals of mathematical statistics},
  pages={148--170},
  year={1961},
  publisher={JSTOR}
}

@article{davis_technology_1989,
  title={Technology acceptance model},
  author={Davis, Fred D and Bagozzi, RP and Warshaw, PR},
  journal={J Manag Sci},
  volume={35},
  number={8},
  pages={982--1003},
  year={1989},
  publisher={Springer}
}

@inproceedings{laugwitz_construction_2008,
  title={Construction and evaluation of a user experience questionnaire},
  author={Laugwitz, Bettina and Held, Theo and Schrepp, Martin},
  booktitle={Symposium of the Austrian HCI and usability engineering group},
  pages={63--76},
  year={2008},
  organization={Springer}
}

@inproceedings{laugwitz_subjektive_2009,
  title={Subjektive Benutzerzufriedenheit quantitativ erfassen: Erfahrungen mit dem User Experience Questionnaire UEQ},
  author={Laugwitz, Bettina and Schubert, Ulf and Ilmberger, Waltraud and Tamm, Nina and Held, Theo and Schrepp, Martin},
  booktitle={Tagungsband UP09},
  pages={220--225},
  year={2009},
  organization={Fraunhofer Verlag}
}

@software{ATLASTI_2025,
  title = {ATLAS.ti},
  author = {{ATLAS.ti Scientific Software Development GmbH}},
  year = {2025},
  organization = {ATLAS.TI},
  version = {25},
  address = {Berlin, Germany},
  url = {https://atlasti.com}
}

@software{whisper,
  author = {OpenAI},
  title = {Whisper: Speech Recognition Model},
  organization = {OpenAI},
  year = {2025},
  howpublished = {https://www.openai.com/whisper},
}

@article{kruskal_use_1952,
  title={Use of ranks in one-criterion variance analysis},
  author={Kruskal, William H and Wallis, W Allen},
  journal={Journal of the American Statistical Association},
  volume={47},
  number={260},
  pages={583--621},
  year={1952},
  publisher={Taylor \& Francis}
}

@article{braun_using_2006,
  title={Using thematic analysis in psychology},
  author={Braun, Virginia and Clarke, Victoria},
  journal={Qualitative research in psychology},
  volume={3},
  number={2},
  pages={77--101},
  year={2006},
  publisher={Taylor \& Francis}
}

@article{fereday_demonstrating_2006,
  title={Demonstrating rigor using thematic analysis: A hybrid approach of inductive and deductive coding and theme development},
  author={Fereday, Jennifer and Muir-Cochrane, Eimear},
  journal={International Journal of Qualitative Methods},
  volume={5},
  number={1},
  pages={80--92},
  year={2006},
  publisher={SAGE Publications}
}

@incollection{harris2011towards,
  title={Towards a theory of learned technological helplessness},
  author={Harris, Joy E},
  booktitle={Encyclopedia of information communication technologies and adult education integration},
  pages={83--101},
  year={2011},
  publisher={IGI Global Scientific Publishing}
}

@article{fichten_higher_2020,
  title={Higher education, information and communication technologies and students with disabilities: An overview of the current situation},
  author={Fichten, Catherine and Olenik-Shemesh, Dorit and Asuncion, Jennison and Jorgensen, Mary and Colwell, Chetz},
  journal={Improving accessible digital practices in higher education: Challenges and new practices for inclusion},
  pages={21--44},
  year={2020},
  publisher={Springer}
}

@inproceedings{suzuki_2003,
author = {Suzuki, Masakazu and Tamari, Fumikazu and Fukuda, Ryoji and Uchida, Seiichi and Kanahori, Toshihiro},
year = {2003},
month = {11},
pages = {95-104},
title = {INFTY: an integrated OCR system for mathematical documents},
booktitle = {Proceedings of the 2003 ACM symposium on Document engineering},
publisher = {ACM},
doi = {10.1145/958220.958239}
}

@inproceedings{karshmer_2002,
  title = {Access to Mathematics by Blind Students},
  booktitle = {Computers Helping People with Special Needs},
  author = {Karshmer, Arthur I. and Bledsoe, Chris},
  editor = {Miesenberger, Klaus and Klaus, Joachim and Zagler, Wolfgang},
  year = 2002,
  pages = {471--476},
  publisher = {Springer Berlin Heidelberg},
  address = {Berlin, Heidelberg},
  isbn = {978-3-540-45491-5}
}

@article{hayes_2024,
  title = {Turning a Blind Eye? {{Removing}} Barriers to Science and Mathematics Education for Students with Visual Impairments},
  shorttitle = {Turning a Blind Eye?},
  author = {Hayes, Cicely and Proulx, Michael J},
  year = 2024,
  month = may,
  journal = {British Journal of Visual Impairment},
  volume = {42},
  number = {2},
  pages = {544--556},
  publisher = {SAGE Publications Ltd},
  issn = {0264-6196},
  doi = {10.1177/02646196221149561},
  urldate = {2025-11-17},
  langid = {english},

}

@article{brinkman_2023,
  title = {Shifting the {{Discourse}} on {{Disability}}: {{Moving}} to an {{Inclusive}}, {{Intersectional Focus}}},
  shorttitle = {Shifting the {{Discourse}} on {{Disability}}},
  author = {Brinkman, Aurora H. and {Rea-Sandin}, Gianna and Lund, Emily M. and Fitzpatrick, Olivia M. and Gusman, Michaela S. and Boness, Cassandra L.},
  year = 2023,
  journal = {The American journal of orthopsychiatry},
  volume = {93},
  number = {1},
  pages = {50--62},
  issn = {0002-9432},
  doi = {10.1037/ort0000653},
  urldate = {2025-11-17},
  pmcid = {PMC9951269},
  pmid = {36265035}
}

@article{chiarella_2020,
  title = {Fieldwork and Disability: An Overview for an Inclusive Experience},
  shorttitle = {Fieldwork and Disability},
  author = {Chiarella, Domenico and Vurro, Grazia},
  year = 2020,
  month = nov,
  journal = {Geological Magazine},
  volume = {157},
  number = {11},
  pages = {1933--1938},
  issn = {0016-7568, 1469-5081},
  doi = {10.1017/S0016756820000928},
  urldate = {2025-11-17},
  langid = {english},
}

@article{stokes_2019,
  title = {Making Geoscience Fieldwork Inclusive and Accessible for Students with Disabilities},
  author = {Stokes, Alison and Feig, Anthony D. and Atchison, Christopher L. and Gilley, Brett},
  year = 2019,
  month = nov,
  journal = {Geosphere},
  volume = {15},
  number = {6},
  pages = {1809--1825},
  issn = {1553-040X},
  doi = {10.1130/GES02006.1},
  urldate = {2025-11-17},
}

@article{goethals_2015,
  title = {Weaving {{Intersectionality}} into {{Disability Studies Research}}: {{Inclusion}}, {{Reflexivity}} and {{Anti-Essentialism}}},
  shorttitle = {Weaving {{Intersectionality}} into {{Disability Studies Research}}},
  author = {Goethals, Tina and Schauwer, Elisabeth De and Hove, Geert Van},
  year = 2015,
  journal = {DiGeSt. Journal of Diversity and Gender Studies},
  volume = {2},
  number = {1-2},
  eprint = {10.11116/jdivegendstud.2.1-2.0075},
  eprinttype = {jstor},
  pages = {75--94},
  publisher = {Leuven University Press},
  issn = {2593-0273},
  doi = {10.11116/jdivegendstud.2.1-2.0075},
  urldate = {2025-11-17},
  
}

@inproceedings{akter_2023,
  title = {``{{If I}}'m Supposed to Be the Facilitator, {{I}} Should Be the Host'': {{Understanding}} the {{Accessibility}} of {{Videoconferencing}} for {{Blind}} and {{Low Vision Meeting Facilitators}}},
  shorttitle = {``{{If I}}'m Supposed to Be the Facilitator, {{I}} Should Be the Host''},
  booktitle = {Proceedings of the 25th {{International ACM SIGACCESS Conference}} on {{Computers}} and {{Accessibility}}},
  author = {Akter, Taslima and Cha, Yoonha and Figueira, Isabela and Branham, Stacy M. and Piper, Anne Marie},
  year = 2023,
  month = oct,
  series = {{{ASSETS}} '23},
  pages = {1--14},
  publisher = {Association for Computing Machinery},
  address = {New York, NY, USA},
  doi = {10.1145/3597638.3608420},
  urldate = {2025-11-20},
  isbn = {979-8-4007-0220-4},

}

@article{das_2019,
  title = {"{{It}} Doesn't Win You Friends": {{Understanding Accessibility}} in {{Collaborative Writing}} for {{People}} with {{Vision Impairments}}},
  shorttitle = {"{{It}} Doesn't Win You Friends"},
  author = {Das, Maitraye and Gergle, Darren and Piper, Anne Marie},
  year = 2019,
  month = nov,
  journal = {Proc. ACM Hum.-Comput. Interact.},
  volume = {3},
  number = {CSCW},
  pages = {191:1--191:26},
  doi = {10.1145/3359293},
  urldate = {2025-11-20},

}

@inproceedings{adnin_2024,
  title = {"{{I}} Look at It as the King of Knowledge": {{How Blind People Use}} and {{Understand Generative AI Tools}}},
  shorttitle = {"{{I}} Look at It as the King of Knowledge"},
  booktitle = {Proceedings of the 26th {{International ACM SIGACCESS Conference}} on {{Computers}} and {{Accessibility}}},
  author = {Adnin, Rudaiba and Das, Maitraye},
  year = 2024,
  month = oct,
  series = {{{ASSETS}} '24},
  pages = {1--14},
  publisher = {Association for Computing Machinery},
  address = {New York, NY, USA},
  doi = {10.1145/3663548.3675631},
  urldate = {2025-11-23},
  isbn = {979-8-4007-0677-6},
 
}

@inproceedings{glazko_2023,
  title = {An {{Autoethnographic Case Study}} of {{Generative Artificial Intelligence}}'s {{Utility}} for {{Accessibility}}},
  booktitle = {Proceedings of the 25th {{International ACM SIGACCESS Conference}} on {{Computers}} and {{Accessibility}}},
  author = {Glazko, Kate S and Yamagami, Momona and Desai, Aashaka and Mack, Kelly Avery and Potluri, Venkatesh and Xu, Xuhai and Mankoff, Jennifer},
  year = 2023,
  month = oct,
  series = {{{ASSETS}} '23},
  pages = {1--8},
  publisher = {Association for Computing Machinery},
  address = {New York, NY, USA},
  doi = {10.1145/3597638.3614548},
  isbn = {979-8-4007-0220-4},

}

@misc{Schonfeld_2019,
  title = {Big {{Deal}}: {{Should Universities Outsource More Core Research Infrastructure}}?},
  author = {Roger C. Schonfeld},
  year = {2019},
  shorttitle = {Big {{Deal}}},
  journal = {Ithaka S+R},
  langid = {english},

}

@misc{Williams_2024,
  title = {Conducting {{Accessibility Research In An Inaccessible Ecosystem}}},
  author = {Michele Williams},
  year = {2024},
  journal = {Smashing Magazine},
  chapter = {General},
  howpublished = {https://www.smashingmagazine.com/2024/04/conducting-accessibility-research-inaccessible-ecosystem/},
  langid = {english}
}

@article{deschamps_2023,
  title = {Better Research Software Tools to Elevate the Rate of Scientific Discovery or Why We Need to Invest in Research Software Engineering},
  author = {Deschamps, Joran and Dalle Nogare, Damian and Jug, Florian},
  year = 2023,
  month = aug,
  journal = {Frontiers in Bioinformatics},
  volume = {3},
  publisher = {Frontiers},
  issn = {2673-7647},
  doi = {10.3389/fbinf.2023.1255159},
  urldate = {2025-11-23},
  langid = {english},
 
}

@article{marathe_2025,
  title = {The {{Accessibility Paradox}}: {{How Blind}} and {{Low Vision Employees Experience}} and {{Negotiate Accessibility}} in the {{Technology Industry}}},
  shorttitle = {The {{Accessibility Paradox}}},
  author = {Marathe, Aparajita S and Piper, Anne Marie},
  year = 2025,
  month = oct,
  journal = {Proc. ACM Hum.-Comput. Interact.},
  volume = {9},
  number = {7},
  pages = {CSCW485:1--CSCW485:25},
  doi = {10.1145/3757666},
  urldate = {2025-11-23},

}

@inproceedings{akter_2025,
author = {Akter, Taslima and Marathe, Aparajita S and Gergle, Darren and Piper, Anne Marie},
title = {Beyond Accessibility: Understanding the Ease of Use and Impacts of Digital Collaboration Tools for Blind and Low Vision Workers},
year = {2025},
isbn = {9798400706769},
publisher = {Association for Computing Machinery},
address = {New York, NY, USA},
url = {https://doi.org/10.1145/3663547.3746332},
doi = {10.1145/3663547.3746332},
booktitle = {Proceedings of the 27th International ACM SIGACCESS Conference on Computers and Accessibility},
articleno = {85},
numpages = {17},
location = {
},
series = {ASSETS '25}
}

@techreport{story_universal_design_1998,
  title = {The {{Universal Design File}}: {{Designing}} for {{People}} of {{All Ages}} and {{Abilities}}. {{Revised Edition}}},
  shorttitle = {The {{Universal Design File}}},
  author = {Story, Molly Follette and Mueller, James L. and Mace, Ronald L.},
  year = 1998,
  institution = {Center for Universal Design, NC State University, Box 8613, Raleigh, NC 27695-8613 (\$24)},
  urldate = {2025-11-25},
  langid = {english},
}

\appendix

\section{Survey Instrument}
\label{sec:survey-instrument}

\subsection{Section 1: Research Tool Usage: Literature Review}

\begin{enumerate}
    \item \textit{[TAM - Actual System Use (ATU)]} Which literature search tools do you use regularly? (Select all that apply.)
    \begin{itemize}
        \item Google Scholar
        \item PubMed
        \item Scopus
        \item Web of Science
        \item JSTOR
        \item ProQuest databases
        \item Other (please specify): \underline{\hspace{3cm}}
        \item I primarily rely on others for literature review (please elaborate – what is the scope of this assistance?): \underline{\hspace{3cm}}
        \item I do not use literature search tools.
    \end{itemize}

    \item \textit{[TAM - ATU]} Which literature management tools do you use regularly? (Select all that apply.)
    \begin{itemize}
        \item Zotero
        \item BibLaTeX
        \item Mendeley
        \item EndNote
        \item RefWorks
        \item F1000Workspace
        \item Other (please specify): \underline{\hspace{3cm}}
        \item I primarily rely on others for literature management (please elaborate – what is the scope of this assistance?): \underline{\hspace{3cm}}
        \item I do not use literature management tools.
    \end{itemize}

    \item \textit{[Learnability] [AT - Subject's perception of Instrument / TAM - Perceived Ease of Use (PEU)]} How easy or difficult is it for you to learn and operate the literature search tools you typically use?
    \begin{itemize}
        \item Very difficult
        \item Difficult
        \item Neutral
        \item Easy
        \item Very easy
    \end{itemize}

    \item \textit{[Usability] [AT - Subject's experience with Instrument / UEQ - Pragmatic Quality]} When using the literature search tools you have access to, how efficiently can you accomplish what you want to do for your literature review?
    \begin{itemize}
        \item Very inefficiently
        \item Inefficiently
        \item Neutral
        \item Efficiently
        \item Very efficiently
    \end{itemize}

    \item \textit{[Satisfaction] [TAM - Perceived Usefulness (PU)]} Overall, how useful are your literature search tools for achieving your literature review objectives?
    \begin{itemize}
        \item Not at all useful
        \item Slightly useful
        \item Moderately useful
        \item Very useful
        \item Extremely useful
    \end{itemize}

    \item \textit{[Learnability] [AT - Subject's perception of Instrument / TAM - Perceived Ease of Use (PEU)]} How easy or difficult is it for you to learn and operate the literature management tools you typically use?
    \begin{itemize}
        \item Very difficult
        \item Difficult
        \item Neutral
        \item Easy
        \item Very easy
    \end{itemize}

    \item \textit{[Usability] [AT - Subject's experience with Instrument / UEQ - Pragmatic Quality]} When using the literature management tools you have access to, how efficiently can you accomplish what you want to do for your literature review?
    \begin{itemize}
        \item Very inefficiently
        \item Inefficiently
        \item Neutral
        \item Efficiently
        \item Very efficiently
    \end{itemize}

    \item \textit{[Satisfaction] [TAM - Perceived Usefulness (PU)]} Overall, how useful are your literature management tools for achieving your literature review objectives?
    \begin{itemize}
        \item Not at all useful
        \item Slightly useful
        \item Moderately useful
        \item Very useful
        \item Extremely useful
    \end{itemize}
\end{enumerate}

\subsection{Section 2: Research Tool Usage: Data Collection}

\begin{enumerate}
    \item \textit{[AT - Instrument / TAM - ATU]} For quantitative data collection, which tools do you use? (Select all that apply)
    \begin{itemize}
        \item Qualtrics
        \item SurveyMonkey
        \item Google Forms
        \item REDCap
        \item SPSS
        \item Excel/Google Sheets for data entry
        \item Custom online forms
        \item Other (please specify): \underline{\hspace{3cm}}
        \item I primarily rely on others for quantitative data collection tasks (please elaborate – what is the scope of this assistance?): \underline{\hspace{3cm}}
        \item I don't conduct quantitative research.
    \end{itemize}

    \item \textit{[AT - Instrument / TAM - ATU]} For qualitative data collection, which tools do you use? (Select all that apply)
    \begin{itemize}
        \item Zoom/Teams recording
        \item Otter.ai or similar transcription services
        \item Digital voice recorders
        \item Interview transcription software
        \item Field note apps
        \item Observation recording tools
        \item Other (please specify): \underline{\hspace{3cm}}
        \item I primarily rely on others for qualitative data analysis tasks (please elaborate – what is the scope of this assistance?): \underline{\hspace{3cm}}
        \item I don't conduct qualitative research.
    \end{itemize}

    \item \textit{[Learnability] [AT - Subject's perception of Instrument / TAM - Perceived Ease of Use (PEU)]} How easy or difficult is it for you to learn and operate the quantitative data collection tools you typically use?
    \begin{itemize}
        \item Very difficult
        \item Difficult
        \item Neutral
        \item Easy
        \item Very easy
    \end{itemize}

    \item \textit{[Usability] [AT - Subject's experience with Instrument / UEQ - Pragmatic Quality]} When using the quantitative data collection tools you have access to, how efficiently can you accomplish what you want to do for your quantitative data collection?
    \begin{itemize}
        \item Very inefficiently
        \item Inefficiently
        \item Neutral
        \item Efficiently
        \item Very efficiently
    \end{itemize}

    \item \textit{[Satisfaction] [AT - Subject's experience with Instrument / UEQ - Hedonic Quality \& SCT - Outcome Expectancy]} Reflecting on your overall experiences, using quantitative data collection tools is generally:
    \begin{itemize}
        \item Very frustrating
        \item Frustrating
        \item Neutral
        \item Satisfying
        \item Very satisfying
    \end{itemize}

    \item \textit{[Learnability] [AT - Subject's perception of Instrument / TAM - Perceived Ease of Use (PEU)]} How easy or difficult is it for you to learn and operate the qualitative data collection tools you typically use?
    \begin{itemize}
        \item Very difficult
        \item Difficult
        \item Neutral
        \item Easy
        \item Very easy
    \end{itemize}

    \item \textit{[Usability] [AT - Subject's experience with Instrument / UEQ - Pragmatic Quality]} When using the qualitative data collection tools you have access to, how efficiently can you accomplish what you want to do for your qualitative data collection?
    \begin{itemize}
        \item Very inefficiently
        \item Inefficiently
        \item Neutral
        \item Efficiently
        \item Very efficiently
    \end{itemize}

    \item \textit{[Satisfaction] [AT - Subject's experience with Instrument / UEQ - Hedonic Quality \& SCT - Outcome Expectancy]} Reflecting on your overall experiences, using qualitative data collection tools is generally:
    \begin{itemize}
        \item Very frustrating
        \item Frustrating
        \item Neutral
        \item Satisfying
        \item Very satisfying
    \end{itemize}
\end{enumerate}

\subsection{Section 3: Research Tool Usage: Data Analysis}
\begin{enumerate}
    \item \textit{[AT - Instrument / TAM - ATU]} What quantitative data analysis tools do you use? (Select all that apply.)
    \begin{itemize}
        \item R/RStudio
        \item SPSS
        \item SAS
        \item Stata
        \item Python
        \item Microsoft Excel/Google Sheets
        \item Other (please specify): \underline{\hspace{3cm}}
        \item I primarily rely on others for quantitative data analysis tasks (please elaborate – what is the scope of this assistance?): \underline{\hspace{3cm}}
        \item I do not use any quantitative data analysis tools.
    \end{itemize}

    \item \textit{[AT - Instrument / TAM - ATU]} What qualitative data analysis tools do you use? (Select all that apply.)
    \begin{itemize}
        \item Microsoft Excel/Google Sheets
        \item QualCoder
        \item NVivo
        \item ATLAS.ti
        \item MAXQDA
        \item Dedoose
        \item Microsoft Word/Google Docs
        \item Other (please specify): \underline{\hspace{3cm}}
        \item I primarily rely on others for qualitative data analysis tasks (please elaborate): \underline{\hspace{3cm}}
        \item I do not use any qualitative data analysis tools.
    \end{itemize}

    \item \textit{[Learnability] [AT - Subject's perception of Instrument / TAM - Perceived Ease of Use (PEU)]} How easy or difficult is it for you to learn and operate the quantitative data analysis tools you typically use?
    \begin{itemize}
        \item Very difficult
        \item Difficult
        \item Neutral
        \item Easy
        \item Very easy
    \end{itemize}

    \item \textit{[Usability] [AT - Subject's experience with Instrument / UEQ - Pragmatic Quality]} When using the quantitative data collection tools you have access to, how efficiently can you accomplish what you want to do for your quantitative data collection?
    \begin{itemize}
        \item Very inefficiently
        \item Inefficiently
        \item Neutral
        \item Efficiently
        \item Very efficiently
    \end{itemize}

    \item \textit{[Satisfaction] [AT - Subject's experience with Instrument / UEQ - Hedonic Quality \& SCT - Outcome Expectancy]} Reflecting on your overall experiences, using quantitative data analysis tools is generally:
    \begin{itemize}
        \item Very frustrating
        \item Frustrating
        \item Neutral
        \item Satisfying
        \item Very satisfying
    \end{itemize}

    \item \textit{[Learnability] [AT - Subject's perception of Instrument / TAM - Perceived Ease of Use (PEU)]} How easy or difficult is it for you to learn and operate the qualitative data analysis tools you typically use?
    \begin{itemize}
        \item Very difficult
        \item Difficult
        \item Neutral
        \item Easy
        \item Very easy
    \end{itemize}

    \item \textit{[Usability] [AT - Subject's experience with Instrument / UEQ - Pragmatic Quality]} When using the qualitative data collection tools you have access to, how efficiently can you accomplish what you want to do for your quantitative data collection?
    \begin{itemize}
        \item Very inefficiently
        \item Inefficiently
        \item Neutral
        \item Efficiently
        \item Very efficiently
    \end{itemize}

    \item \textit{[Satisfaction] [AT - Subject's experience with Instrument / UEQ - Hedonic Quality \& SCT - Outcome Expectancy]} Reflecting on your overall experiences, using qualitative data analysis tools is generally:
    \begin{itemize}
        \item Very frustrating
        \item Frustrating
        \item Neutral
        \item Satisfying
        \item Very satisfying
    \end{itemize}
\end{enumerate}

\subsection{Section 4: Research Tool Usage: Manuscript Writing}
\begin{enumerate}
    \item \textit{[AT - Instrument / TAM - ATU]} What manuscript writing tools do you use? (Select all that apply.)
    \begin{itemize}
        \item Microsoft Word
        \item Overleaf (online LaTeX editor)
        \item Google Docs
        \item Writefull
        \item Grammarly
        \item Other (please specify): \underline{\hspace{3cm}}
        \item I primarily rely on others for manuscript writing tasks (please elaborate – what is the scope of this assistance?): \underline{\hspace{3cm}}
        \item I do not use any manuscript writing tools.
    \end{itemize}

    \item \textit{[Learnability] [AT - Subject's perception of Instrument / TAM - Perceived Ease of Use (PEU)]} How easy or difficult is it for you to learn and operate the manuscript writing tools you typically use?
    \begin{itemize}
        \item Very difficult
        \item Difficult
        \item Neutral
        \item Easy
        \item Very easy
    \end{itemize}

    \item \textit{[Usability] [AT - Subject's experience with Instrument / UEQ - Pragmatic Quality]} When using the manuscript writing tools you have access to, how efficiently can you accomplish what you want to do for your quantitative data collection?
    \begin{itemize}
        \item Very inefficiently
        \item Inefficiently
        \item Neutral
        \item Efficiently
        \item Very efficiently
    \end{itemize}

    \item \textit{[Satisfaction] [AT - Subject's experience with Instrument / UEQ - Hedonic Quality \& SCT - Outcome Expectancy]} Reflecting on your overall experiences, using manuscript writing tools is generally:
    \begin{itemize}
        \item Very frustrating
        \item Frustrating
        \item Neutral
        \item Satisfying
        \item Very satisfying
    \end{itemize}
\end{enumerate}

\subsection{Section 5: Accessibility Challenges}
\begin{enumerate}
    \item \textit{[AT - Contradiction Identification (Subject-Instrument-Object)]} Which research stage presents the most accessibility challenges for you? (Rank from 1 = most challenging to 5 = least challenging)
    \begin{itemize}
        \item Literature review and reference management
        \item Data collection (surveys, interviews, observations)
        \item Data analysis and visualization
        \item Writing and manuscript preparation
        \item Presentation and dissemination
    \end{itemize}

    \item \textit{[AT - Instrument Properties (Source of Contradiction)]} Which specific research tools are most problematic for accessibility? (Select up to 5)
    \begin{itemize}
        \item Statistical software interfaces
        \item Survey creation platforms
        \item Data visualization tools
        \item Reference managers
        \item Collaboration platforms
        \item Database search interfaces
        \item Transcription software
        \item Presentation software
        \item Other (please specify): \underline{\hspace{3cm}}
    \end{itemize}

    \item \textit{[AT - Instrument Properties (Source of Contradiction) / SCT - Environmental Barrier]} What are the most common accessibility barriers you encounter during the literature review stage?
    \begin{itemize}
        \item[] [Open text]
    \end{itemize}

    \item \textit{[AT - Instrument Properties (Source of Contradiction) / SCT - Environmental Barrier]} What are the most common accessibility barriers you encounter during the data collection stage?
    \begin{itemize}
        \item[] [Open text]
    \end{itemize}

    \item \textit{[AT - Instrument Properties (Source of Contradiction) / SCT - Environmental Barrier]} What are the most common accessibility barriers you encounter during the data analysis stage?
    \begin{itemize}
        \item[] [Open text]
    \end{itemize}

    \item \textit{[AT - Contradiction Impact on Behavior / TAM - Behavioral Intention]} To what extent do accessibility barriers in research tools during the literature review stage lead you to avoid using them, even if they are standard in your field (community/rules)?
    \begin{itemize}
        \item Not at all
        \item Slightly
        \item Moderately
        \item To a great extent
        \item Completely
    \end{itemize}

    \item \textit{[AT - Contradiction Impact on Behavior / TAM - Behavioral Intention]} To what extent do accessibility barriers in research tools during the data collection stage lead you to avoid using them, even if they are standard in your field (community/rules)?
    \begin{itemize}
        \item Not at all
        \item Slightly
        \item Moderately
        \item To a great extent
        \item Completely
    \end{itemize}

    \item \textit{[AT - Contradiction Impact on Behavior / TAM - Behavioral Intention]} To what extent do accessibility barriers in research tools during the data analysis stage lead you to avoid using them, even if they are standard in your field (community/rules)?
    \begin{itemize}
        \item Not at all
        \item Slightly
        \item Moderately
        \item To a great extent
        \item Completely
    \end{itemize}

    \item \textit{[AT - Contradiction Impact on Behavior / TAM - Behavioral Intention]} To what extent do accessibility barriers in research tools during the manuscript writing stage lead you to avoid using them, even if they are standard in your field (community/rules)?
    \begin{itemize}
        \item Not at all
        \item Slightly
        \item Moderately
        \item To a great extent
        \item Completely
    \end{itemize}

    \item \textit{[AT - Contradiction Example (Subject-Instrument-Object)]} Describe your most significant accessibility challenge with a specific research tool and how it impacted your ability to perform a research task.
    \begin{itemize}
        \item[] [Open text]
    \end{itemize}
\end{enumerate}

\subsection{Section 6: Strategies and Workarounds}
\begin{enumerate}
    \item \textit{[AT - Subject Agency \& Strategy / SCT - Behavioral Coping]} When you encounter an inaccessible research tool, what do you typically do? (Select all that apply.)
    \begin{itemize}
        \item Find an alternative accessible tool
        \item Ask a sighted colleague for help
        \item Use assistive technology workarounds
        \item Contact the vendor for accessibility support
        \item Modify my research approach/methodology
        \item Work around the tool using other software
        \item Abandon that particular task/tool
        \item Other (please specify): \underline{\hspace{3cm}}
    \end{itemize}

    \item \textit{[AT - Subject Beliefs / SCT - Self-Efficacy]} How confident are you in your ability to find or develop ways to continue your research work when faced with inaccessible tools?
    \begin{itemize}
        \item Not at all confident
        \item Slightly confident
        \item Moderately confident
        \item Very confident
        \item Extremely confident
    \end{itemize}

    \item \textit{[AT - Subject Innovation (Resolving Contradiction)]} Please describe a specific creative workaround or strategy you've developed or used to overcome an accessibility barrier in a research tool to achieve a research goal.
    \begin{itemize}
        \item[] [Open text]
    \end{itemize}

    \item \textit{[AT - Rules \& Community Influence on Subject Response]} To what extent do you feel the official or prescribed methods/tools (rules/community norms) for research tasks in your environment accommodate your accessibility needs when selecting or using tools?
    \begin{itemize}
        \item Not at all
        \item Slightly
        \item Moderately
        \item To a great extent
        \item Completely accommodated
    \end{itemize}
\end{enumerate}

\subsection{Section 7: Collaborative Experiences}
\begin{enumerate}
    \item \textit{[AT - Community Interaction \& Division of Labor / SCT - Behavioral Strategy]} When research tools are inaccessible to you, how do you typically manage collaborative work (division of labor) with colleagues? (Select all that apply.)
    \begin{itemize}
        \item Request colleagues to operate tools for me
        \item Use screen sharing with colleague assistance
        \item Divide tasks based on tool accessibility
        \item Work with separate accessible tools and merge results
        \item Avoid collaborative projects requiring inaccessible tools
        \item Ask for accessible file formats/outputs
        \item Other (please specify): \underline{\hspace{3cm}}
    \end{itemize}

    \item \textit{[AT - Community Support / SCT - Environmental Factor (Social Support)]} In your experience, how supportive are your colleagues in adapting workflows or assisting when research tools are inaccessible to you?
    \begin{itemize}
        \item Not at all supportive
        \item Slightly supportive
        \item Moderately supportive
        \item Very supportive
        \item Extremely supportive
    \end{itemize}

    \item \textit{[AT - Rules \& Division of Labor in Community]} How is the choice of research tools typically decided in your collaborative projects (e.g., PI decides, group consensus, individual choice, depends on existing lab tools)?
    \begin{itemize}
        \item[] (Open text)
    \end{itemize}

    \item \textit{[AT - Rules \& Community Norms]} Are accessibility considerations for all team members typically part of the decision-making process when selecting tools in your collaborative projects?
    \begin{itemize}
        \item Never
        \item Rarely
        \item Sometimes
        \item Often
        \item Always
    \end{itemize}

    \item \textit{[AT - Contradiction within Community/Division of Labor]} Describe a specific collaboration challenge related to tool accessibility and how you and your colleagues (community) addressed it (or were unable to address it), and how this affected the shared work.
    \begin{itemize}
        \item[] [Open text]
    \end{itemize}
\end{enumerate}

\subsection{Section 8: Impact on Research Work/Productivity}
\begin{enumerate}
    \item \textit{[AT - Outcome / SCT - Experienced Outcomes]} How do accessibility barriers in research tools impact your work? (Select all that apply.)
    \begin{itemize}
        \item Significantly increase time needed for tasks
        \item Limit my independence in research
        \item Influence my choice of research topics/methods
        \item Reduce opportunities for collaboration
        \item Require additional training or support
        \item Increase research-related expenses
        \item Affect quality of research outputs
        \item No significant impact
        \item Other (please specify): \underline{\hspace{3cm}}
    \end{itemize}

    \item \textit{[AT - Outcome (related to Rules \& Community)]} How do tool accessibility issues affect your ability to meet the expectations or timelines set by your institution or research community?
    \begin{itemize}
        \item No effect
        \item Minor negative effect
        \item Moderate negative effect
        \item Significant negative effect
        \item Profound negative effect / makes it impossible to meet them
    \end{itemize}

    \item \textit{[SCT - Outcome Expectancies / Experienced Outcomes (Career Level)]} Accessibility barriers have influenced major decisions about my research career.
    \begin{itemize}
        \item Strongly disagree
        \item Disagree
        \item Neutral
        \item Agree
        \item Strongly agree
    \end{itemize}

    \item \textit{[AT - Elaboration on Outcome]} If you selected 'Agree' or 'Strongly Agree' to the previous question, please elaborate briefly on how these barriers have influenced your career decisions.
    \begin{itemize}
        \item[] [Open text]
    \end{itemize}
\end{enumerate}

\subsection{Section 9: Community Support}
\begin{enumerate}
    \item \textit{[AT - Community \& Rules (Knowledge Sharing)]} Beyond official documentation or formal training, how have you learned effective strategies or workarounds for research tools? (Select all that apply.)
    \begin{itemize}
        \item From a mentor or advisor
        \item From colleagues in my lab/team
        \item From peers in my field but outside my institution
        \item Through online forums, mailing lists, or social media groups for BLV individuals
        \item I have had to discover most of them on my own through trial and error
        \item My institution provides specific, helpful training on accessible tool use
        \item Other (please specify): \underline{\hspace{3cm}}
        \item I have not found effective strategies or workarounds
    \end{itemize}

    \item \textit{[AT - Community Influence on Outcome]} When you face challenges with a tool's usability or accessibility, how much does support from your research community help mitigate the negative impact on your work?
    \begin{itemize}
        \item Not at all; I have to handle it on my own.
        \item Slightly, the support provides minor help.
        \item Moderately; the support is helpful, but doesn't solve the core problem.
        \item A great deal; the support is essential for me to overcome the challenge.
        \item I do not receive support from my community for these issues.
    \end{itemize}

    \item \textit{[AT - Community Influence on Outcome]} Thinking about a time you had to learn a difficult new tool, how did your community environment affect its learnability for you?
    \begin{itemize}
        \item The community made it much harder to learn (e.g., by being unsupportive or creating pressure).
        \item The community did not affect my learning experience.
        \item The community made it slightly easier to learn (e.g., by answering a few questions).
        \item The community made it much easier to learn (e.g., through mentorship, shared resources, or collaborative learning).
        \item This scenario does not apply to me.
        \item[] Please elaborate on your selection. [Open text]
    \end{itemize}

    \item \textit{[AT - Outcome (related to Community)]} To what extent does the level of support from your research community affect your overall satisfaction with your work as a researcher?
    \begin{itemize}
        \item It greatly decreases my satisfaction.
        \item It somewhat decreases my satisfaction.
        \item It does not affect my satisfaction.
        \item It somewhat increases my satisfaction.
        \item It greatly increases my satisfaction.
    \end{itemize}

    \item \textit{[AT - Community (Institutional Level) \& Rules]} How effective have official university/institutional resources (e.g., Disability Resource Center, IT Accessibility Office, departmental funding) been in resolving accessibility challenges with specific research tools for you?
    \begin{itemize}
        \item Very ineffective
        \item Ineffective
        \item Neutral
        \item Effective
        \item Very effective
    \end{itemize}

    \item \textit{[AT - Community Agency \& Advocacy]} To your knowledge, has your research community (e.g., lab, department, professional organization) ever collectively advocated for more accessible research tools from a vendor or your institution?
    \begin{itemize}
        \item Yes, and it was successful.
        \item Yes, but it was unsuccessful.
        \item No, to my knowledge, this has not happened.
        \item I don't know.
    \end{itemize}

    \item \textit{[AT - Contradiction Example (involving Community)]} Please describe a time when support from your wider research community (beyond direct collaborators) made a significant difference—either positive or negative—in your ability to conduct your research.
    \begin{itemize}
        \item[] [Open text]
    \end{itemize}
\end{enumerate}

\subsection{Section 10: Demographics and Follow-Up Interview}
\begin{enumerate}
    \item What is your age in years?
    \begin{itemize}
        \item 18-24
        \item 25-34
        \item 35-44
        \item 45-54
        \item 55-64
        \item 65 or older
    \end{itemize}

    \item What is your gender?
    \begin{itemize}
        \item Woman
        \item Man
        \item Non-binary
        \item Prefer to self-describe
        \item Prefer not to say
    \end{itemize}

    \item What is your level of visual acuity?
    \begin{itemize}
        \item Totally blind (no light or shape perception)
        \item Legally blind, with both light and shape perception
        \item Legally blind, with only light perception
        \item Legally blind, with only shape perception
        \item Legally blind, central vision loss
        \item Legally blind, peripheral vision loss
        \item Legally blind, tunnel vision
        \item Legally blind, blurry vision
        \item Legally blind, fluctuating vision
        \item Legally blind, partial sight
        \item Prefer to self-describe: \underline{\hspace{3cm}}
    \end{itemize}

    \item \textit{[AT - Subject]} In what context do you conduct research?
    \begin{itemize}
        \item Academia (professor)
        \item Academia (post-doctoral researcher)
        \item Graduate student (PhD)
        \item Graduate student (Master’s)
        \item Academia (undergraduate student)
        \item Industry (research and development; R\&D)
        \item Independent researcher (outside of academia or industry)
        \item Other (please specify): \underline{\hspace{3cm}}
    \end{itemize}

    \item \textit{[AT - Subject]} What is your primary research discipline or field?
    \begin{itemize}
        \item Arts \& Humanities (e.g., Classics, English, History, Languages, Philosophy, Visual/Performing Arts)
        \item Social Sciences (e.g., Anthropology, Communication, Economics, Education, Political Science, Psychology, Sociology)
        \item Life \& Health Sciences (e.g., Biochemistry, Biology, Biomedical Sciences, Genetics, Medicine, Nursing, Public Health)
        \item Physical Sciences (e.g., Astronomy, Chemistry, Earth Sciences, Geology, Oceanography, Physics)
        \item Formal Sciences (e.g., Mathematics, Statistics, Biostatistics, Computer Science, AI/Machine Learning)
        \item Engineering \& Technology (e.g., Aerospace, Biomedical, Chemical, Civil, Electrical, Mechanical, Software Engineering)
        \item Interdisciplinary Studies / Other (please specify): \underline{\hspace{3cm}}
    \end{itemize}

    \item How many years of experience do you have in research?
    \begin{itemize}
        \item Less than 2 years
        \item 2-5 years
        \item 6-10 years
        \item 11-20 years
        \item More than 20 years
    \end{itemize}

    \item Do you use assistive technologies for accessing computers or digital information? (Select all that apply.)
    \begin{itemize}
        \item Screen reader (e.g., JAWS, NVDA, VoiceOver)
        \item Screen magnification software (e.g., ZoomText, Windows Magnifier)
        \item Braille display
        \item Large print
        \item High contrast mode/custom color schemes
        \item Voice input/dictation software (e.g., Dragon NaturallySpeaking)
        \item Other (please specify): \underline{\hspace{3cm}}
        \item I do not use assistive technologies for these tasks
    \end{itemize}

    \item Are you interested in being contacted about a paid follow-up interview with a graduate student researcher from the research team?
    \begin{itemize}
        \item No
        \item Yes (If yes, please provide your email address. It will only be used for this purpose): \underline{\hspace{3cm}}
    \end{itemize}
\end{enumerate}

\section{Interview Script}
\label{sec:interview-script}

\subsection*{Introduction}

\begin{itemize}
    \item \textbf{[Interviewer]} Hi, how are you today? Thank you for taking the time to speak with me today! Before we get started, I wanted to give you a quick refresher of the study just in case. The purpose of this study is to gain a better understanding of the experiences of blind and low-vision (BLV) researchers, particularly their daily workflows and processes as researchers in their respective fields. More specifically, we're interested in learning more about the kinds of research tools you use during the various stages of the research process, your experiences with their accessibility, their impact on your collaboration with fellow researchers and practitioners, and what your broader thoughts are on accessible research tooling.
    \item \textbf{[Interviewer]} This interview should take about 45 minutes to an hour. I'll be asking you about your research background, the tools you use, challenges you face, and strategies you've developed. Please feel free to share as much detail as you're comfortable with---your experiences are really valuable for this research.
    \item \textbf{[Interviewer]} Before we begin, do you have any questions about the study or the interview process? \ldots Great! Let's review the informed consent form.
\end{itemize}

\subsubsection*{Share and review the informed consent form.}

\begin{itemize}
    \item \textbf{[Interviewer]} This form covers information related to the study and to today's interview. During the interview, I'll ask you several questions about your feelings on your research workflow and your thoughts on the accessibility of research tools. As I mentioned, the interview will last between 45 minutes and an hour, and you'll be compensated with a \$30 Amazon e-gift card as a token of our appreciation. You do not have to answer any question that you are not comfortable answering, and you are free to end the interview at any time.
    \item \textbf{[Interviewer]} With your permission, I'd like to record this conversation so I can focus on our discussion rather than taking notes. The recording will only be used for transcription purposes and will be kept confidential. Is that okay with you?
    \item \textbf{[Interviewer]} Great! Let's get started.
\end{itemize}

\subsection*{Part I: Research Background and Context (8--10 minutes)}

\begin{itemize}
    \item \textbf{[Interviewer]} First, I'd like to learn a bit about your research background. Can you tell me about your current research role (e.g., graduate student, professor, industry, etc.) and what field you work in?
    \item \textbf{[Interviewer]} How would you describe your visual acuity, and what assistive technologies do you primarily rely on for your work?
    \item \textbf{[Interviewer]} How long have you been conducting research, and what drew you to your particular area of study?
    \item \textbf{[Interviewer]} Can you walk me through what a typical research project looks like for you from start to finish? What are the main stages you go through?
    \item \textbf{[Interviewer]} What types of research do you primarily conduct---qualitative, quantitative, or a mix of both?
    \item \textbf{[Interviewer]} Has your approach to research changed over time as you've gained experience or as technology has evolved?
    \item \textbf{[Probes]} What motivated those changes? How has technology evolution affected your work?
\end{itemize}

\subsection*{Part II: Research Tool Usage and Workflows (12--15 minutes; RQ1)}

\begin{itemize}
    \item \textbf{[Interviewer]} Now, I'd like to dive into the specific tools you use. Let's start with literature review---what's your process for finding and managing research literature?
    \item \textbf{[Interviewer]} What tools do you use for literature search, and how do you typically organize and manage your references?
    \item \textbf{[Probes]} Which databases? How do you handle PDFs? Reference management workflow?
    \item \textbf{[Interviewer]} Moving to data collection---can you walk me through how you typically collect data for your research?
    \item \textbf{[Interviewer]} What specific tools or platforms do you use for data collection, and how do you set up your data collection process?
    \item \textbf{[Probes]} If interviews---recording setup? If surveys---platform choice? If observations---note-taking methods?
    \item \textbf{[Interviewer]} For data analysis, what's your typical workflow? What tools do you rely on most heavily?
    \item \textbf{[Interviewer]} How do you approach data visualization or presenting your findings? What tools work best for you in this stage?
    \item \textbf{[Probes]} How do you handle graphs/charts? What about tables and statistical outputs?
    \item \textbf{[Interviewer]} Finally, for writing and dissemination, what tools do you use to write up your research and share it with others?
\end{itemize}

\subsection*{Part III: Strategies and Workarounds (8--10 minutes; RQ3)}

\begin{itemize}
    \item \textbf{[Interviewer]} I'm curious about the creative solutions you've developed. Can you tell me about a specific workaround or strategy you've created to deal with an inaccessible tool?
    \item \textbf{[Interviewer]} How do you typically approach learning to use a new research tool? What's your process for evaluating its accessibility?
    \item \textbf{[Interviewer]} When you encounter an accessibility barrier, what's your decision-making process? Do you try to work around it, find alternatives, or something else?
    \item \textbf{[Interviewer]} Have you ever contacted software vendors or developers about accessibility issues? What was that experience like?
    \item \textbf{[Interviewer]} Are there any general strategies or principles that guide how you set up your research workflow to minimize accessibility challenges?
    \item \textbf{[Probes]} File organization? Tool selection criteria? Backup methods?
    \item \textbf{[Interviewer]} What resources or support systems have been most helpful in developing these strategies?
\end{itemize}

\subsection*{Part IV: Collaboration and Social Aspects (8--10 minutes; RQ4)}

\begin{itemize}
    \item \textbf{[Interviewer]} Let's talk about collaboration. How do accessibility challenges affect your ability to work with colleagues or research teams?
    \item \textbf{[Probe]} Can you describe a specific collaborative project where tool accessibility became an issue? How did you and your team handle it?
    \item \textbf{[Interviewer]} How do you typically communicate your accessibility needs to collaborators, especially when starting new projects?
    \item \textbf{[Interviewer]} What has been your experience with colleagues' awareness and responsiveness to accessibility issues?
    \item \textbf{[Probes]} Any particularly positive or negative experiences? How do you educate colleagues?
    \item \textbf{[Interviewer]} How do you handle situations where the rest of your team wants to use a tool that's inaccessible to you?
    \item \textbf{[Interviewer]} Do you think accessibility challenges have affected your professional relationships or broader networking opportunities?
\end{itemize}

\subsection*{Part V: Recommendations \& Future Vision (8--10 minutes; RQ5)}

\begin{itemize}
    \item \textbf{[Interviewer]} If you could give advice to software developers creating research tools, what would be your top three recommendations for improving accessibility?
    \item \textbf{[Interviewer]} What about advice for institutions or research organizations---what could they do to better support BLV researchers?
    \item \textbf{[Interviewer]} Do you have experience with using AI or AI-assisted tools as a part of your research workflow? If not, do you plan to use AI in the future, and if so, what features do you hope these tools will have for you?
    \item \textbf{[Probe]} Asking detailed questions about features they want to have.
    \item \textbf{[Probe]} What do you think is not readily accessible to you now, but can become accessible with AI as a part of your research process?
    \item \textbf{[Interviewer]} If you could wave a magic wand and make one change to improve research tool accessibility, what would it be?
    \item \textbf{[Interviewer]} What advice would you give to other BLV researchers who might be struggling with accessibility challenges in their work?
    \item \textbf{[Interviewer]} How do you think the landscape of research tool accessibility has changed over your career, and where do you see it heading?
    \item \textbf{[Interviewer]} Is there anything important about your experiences with research tool accessibility that we haven't covered today?
\end{itemize}

\subsection*{Wrap-Up and Final Thoughts}

\begin{itemize}
    \item \textbf{[Interviewer]} We're coming to the end of our time together. Is there anything else you'd like to share about your experiences as a BLV researcher?
    \item \textbf{[Interviewer]} Are there any questions you have for me about this research or how we plan to use these insights?
\end{itemize}

\subsubsection*{Stop the recording.}

\begin{itemize}
    \item \textbf{[Interviewer]} Thank you so much for sharing your experiences and insights with me today. Your perspective is incredibly valuable for this research and for improving accessibility in research tools more broadly.
\end{itemize}

\section{Participant Demographics}
\label{sec:participant-demographics}
\subsection{Survey Participants}
\label{subsec:survey-participants}

\begin{table}[htbp]
\centering
\captionsetup{justification=centering}
\caption{Survey respondent demographics (n=57).}
\label{tab:demographics}
\Description{A summary table of demographics for 57 survey respondents, broken down by percentage across five categories. For Age Range, the largest group is 25-34 at 33.3\%. For Gender, 54.4\% identify as Woman. For Visual Acuity, 39.1\% are totally blind and 34.8\% are legally blind with only light perception. For Research Setting, 50.0\% are graduate students. For Field of Study, 52.4\% are in the Social Sciences.}
\begin{tabular}{@{}lrl@{}}
\toprule
\textbf{Characteristic} & \textbf{\%} \\
\midrule

\textbf{Age Range} & \\
\quad 25--34 & 33.3 \\
\quad 35--44 & 28.1 \\
\quad 45--54 & 15.8 \\
\quad 18--24 & 14.0 \\
\quad 55--64 & 7.0 \\
\quad 65+ & 1.8 \\
\midrule

\textbf{Gender} & & \\
\quad Woman & 54.4 \\
\quad Man & 38.6 \\
\quad Non-binary & 5.3 \\
\quad Prefer not to answer & 1.8 \\
\midrule

\textbf{Visual Acuity} & & \\
\quad  Totally blind (no light or shape perception)  & 39.1 \\
\quad Legally blind, with only light perception: & 34.8 \\
\quad Prefer to self-describe & 13.0 \\
\quad Legally blind, with both light and shape perception & 8.7 \\
\quad Legally blind, partial sight & 4.3 \\
\midrule

\textbf{Research Setting} & & \\
\quad Graduate student (Master's/Ph.D.) & 50.0 \\
\quad  Academia (professor) & 17.9 \\
\quad Independent researcher (outside of academia or industry) & 14.3 \\
\quad Research scientists (academia) & 14 \\
\quad Other & 10.7 \\
\quad Industry (research and development; R\&D) & 7.1 \\
\midrule

\textbf{Field of Study} & & \\
\quad Social Sciences & 52.4 \\
\quad Interdisciplinary Studies & 14.3 \\
\quad Life \& Health Sciences & 14.3 \\
\quad Physical Sciences & 9.5 \\
\quad Formal Sciences & 4.8 \\
\quad Engineering & 4.8 \\

\bottomrule
\end{tabular}
\end{table}

\clearpage
\onecolumn
\subsection{Interview Participants}
\label{subsec:interview-participants}

\begin{table*}[h]
\centering
\captionsetup{justification=centering}
\small
\caption{Demographics of interview participants (n-15). NA entries indicate that participants declined to provide a specific data value.}
\begin{tabular}{@{}lp{2.5cm}p{3cm}cp{3cm}@{}}
\toprule
\textbf{Participant} & \textbf{Discipline} & \textbf{Occupation} & \textbf{Years in Research} & \textbf{Visual Acuity} \\
\midrule
P1 & Economics \& Management & Ph.D. Candidate & 5 & Legally blind, light perception \\
P2 & Italian Literature & Research Associate \& Ph.D. Candidate & 10 & Totally blind \\
P3 & Special Education & Ph.D Candidate & 7 & Legally blind, light perception \\
P4 & Sociology & Research Director & 20 & Totally blind \\
P5 & International Relations & Graduate Student & 2 & Legally blind, blurry vision \\
P6 & Business & Undergraduate Research Assistant & 2 & Low-vision \\
P7 & Public Health & Research Specialist & 7 & Legally blind \\
P8 & Kinesiology & Assistant Professor & 6 & Totally blind, light perception \\
P9 & Law & Lawyer & NA & Legally blind \\
P10 & Library Sciences & Associated Professor & 12 & Low-vision, 20/125 \\
P11 & Queer \& Gender Studies & Graduate Student & 2 & Legally blind, light perception \\
P12 & Music Education & Music Instructor & 2 & Totally blind, light perception \\
P13 & Educational Technology & Non-Profit CEO & 20 & Legally blind \\
P14 & Plasma Physics & Research Director & 40 & Totally blind, light perception \\
P15 & Atmospheric Sciences & Meteorologist & 10 & Legally blind \\
\bottomrule
\end{tabular}
\label{tab:reduced_demographics}
\Description{Table showing demographics of 15 interview participants (P1-P15) with columns for Participant ID, Discipline, Occupation, Years in Research, and Visual Acuity. Disciplines span humanities (Italian Literature, Queer & Gender Studies), social sciences (Sociology, Law, Economics), STEM fields (Plasma Physics, Atmospheric Sciences, Kinesiology), and applied fields (Business, Public Health, Educational Technology). Occupations include Ph.D. candidates (5 participants), graduate students (2), faculty members (2 professors), research specialists and directors (3), and practitioners (lawyer, meteorologist, music instructor, non-profit CEO). Research experience ranges from 2 to 40 years, with one participant's years not reported. Visual acuity varies: 7 participants are legally blind (4 with light perception only), 6 are totally blind (3 with light perception), and 2 have low vision (one specified as 20/125). The table illustrates diversity in academic disciplines and career stages among BLV researchers, with all participants having significant vision loss.}
\end{table*}

\end{document}